\input {epsf}
\documentstyle[prl,aps, multicol,epsf]{revtex}
\begin{document}
\draft
\title{Numerical studies of the phase diagram of layered type II  
superconductors in a magnetic field} 
\author{A.K.Kienappel and M.A.Moore}
\address{Department of Physics, University of Manchester,
Manchester, M13 9PL, United Kingdom.}
\date{\today}
\maketitle

\begin{abstract}
We report on simulations of layered superconductors using the 
Lawrence-Doniach model 
in the framework of the lowest Landau level approximation. 
We find  a first order phase transition with a $B(T)$ dependence
which agrees very well with the experimental
``melting'' line in YBa$_{2}$Cu$_{3}$O$_{7-\delta}$.   
The transition is not associated with vortex lattice melting, but 
separates two vortex liquid states characterised by different 
degrees of short-range crystalline order and different length 
scales of correlations between vortices in different layers.
The transition line ends at a critical end-point at low fields. 
We find the magnetization 
discontinuity and the location of the lower critical magnetic field
to be in good agreement with experiments  
in YBa$_{2}$Cu$_{3}$O$_{7-\delta}$.
Length scales of order parameter correlations parallel and
perpendicular to the magnetic field increase 
exponentially as {\small $1/T$} at low temperatures. The dominant 
relaxation time scales grow roughly exponentially with these
correlation lengths. We find that the first order phase 
transition  persists in the presence of weak random 
point disorder but can be suppressed entirely by strong disorder. 
No vortex glass or Bragg glass state is found in the
presence of disorder. 
The consistency of our numerical results with various 
experimental features in YBa$_{2}$Cu$_{3}$O$_{7-\delta}$, 
including the dependence on anisotropy, and the temperature dependence of
the structure factor at the  Bragg peaks in neutron 
scattering experiments is demonstrated.
\end{abstract}
\begin{multicols}{2}
\narrowtext

\section{Introduction}
\begin{figure}
\centerline{\epsfxsize= 8 cm\epsfbox{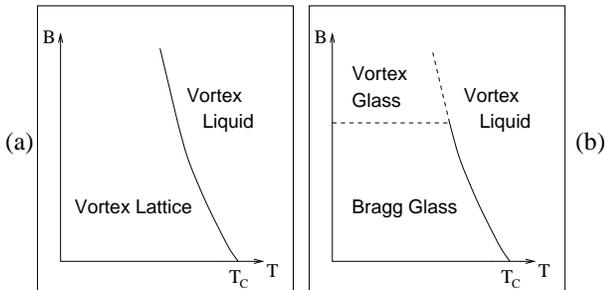}}
  \caption{Popular phase diagrams (a) in the clean case and 
   (b) in the presence of  disorder. Solid and dotted lines mark first 
   order and continuous transitions respectively.}
  \label{fig:compd}
\end{figure}

In the mean-field limit the phase diagram of type II superconductors
has two phases: the normal state and 
the mixed state in which the lines of magnetic flux 
are arranged in a triangular Abrikosov lattice \cite{parks}. 
However, thermal fluctuations destroy  the flux 
lattice near the mean-field transition
line and a flux liquid phase enters the phase diagram
\cite{thouless&ruggeri}.  
As the temperature is reduced the  vortex liquid undergoes a first order
phase transition to what is commonly assumed to be the 
flux lattice state. 
This leads to a phase diagram as shown in Fig.\ \ref{fig:compd}(a). 
The strong belief in first order melting of the 
Abrikosov flux lattice rests on the experimental 
evidence reviewed in Sec.\ \ref{sec:ybcoexp}. 
Much of the analytical work on vortex lattice melting 
relies on the Lindemann criterion, 
which states that melting occurs if the mean fluctuation radius of a 
lattice point around its equilibrium position has reached a 
certain fraction (usually between 0.1 and 0.2) of the lattice
constant \cite{blatterrev}. This criterion is not rigorous 
and does not provide a satisfying thermodynamic melting theory. 
The possibility of a first order phase transition 
due to decoupling of the different layers has also been investigated 
\cite{glazman&koshelev&daemen&bulaevsky}. 
However, a decoupling transition
is mostly expected to occur in addition
to melting, and the lack of experimental evidence for two separate 
phase transitions has lead to a widespread belief that 
either there is no sharp decoupling transition or that it occurs 
simultaneously with flux lattice melting.
Our numerical results suggest a phase diagram which is fundamentally 
different from Fig.\ \ref{fig:compd}(a). It
has a first order phase transition in excellent agreement 
 with the first order
transition line in YBa$_{2}$Cu$_{3}$O$_{7}$ (YBCO) in the $B$-$T$ 
plane (see Sec.\ \ref{sec:numpd}). 
However, this transition is a decoupling transition and 
not associated with vortex lattice melting. 
There is only one phase in the phase diagram: the vortex liquid phase,
and a  vortex lattice exists only at zero temperature. 

Although there is striking experimental evidence 
for a first order phase transition in both 
YBCO and 
Bi$_{2}$Sr$_{2}$CaCu$_{2}$O$_{8}$ (BSCCO), there are certain
features of the experimental data that are not explained by 
the standard vortex lattice melting picture, most importantly  
the loss of first order
behavior along the transition line at high (both for YBCO and BSCCO)
and low (YBCO only)  fields. 
Note that an end of the first order phase transition line
at a critical end-point
is not possible for a vortex lattice melting line,
because the phase boundary separates phases of different symmetry.
Our first order transition is not associated
with any symmetry breaking. Thus the existence
of a low field critical end-point should 
be expected and  is directly observed in the simulation. 

In the framework
of a vortex lattice melting picture the disappearance of the 
first order melting line can be explained by the presence of a 
tricritical point
where the first order transition changes to a continuous one. 
Such behavior is commonly assumed to occur as an effect of sample disorder,
which is to a certain degree present even in the best crystals.   
However, there is no wide consensus on the phase diagram 
in the presence of disorder. 
The three most important categories to distinguish are the 
disordered liquid, vortex glass and Bragg glass scenarios. 
For the first case there is no phase which is thermodynamically 
distinct from a vortex 
liquid and thus no thermodynamic phase transition. 
However, a fairly sharp crossover from fast to slow dynamics may occur
within the vortex liquid state. 
The vortex glass scenario presented in detail
in Ref.\ \cite{2fisher&huse} relies on analogy to spin glass behavior. 
The vortex liquid is expected to freeze via a continuous transition
to a vortex glass state characterized by short-range crystalline 
correlations but long-range phase correlations.
Such a vortex glass phase would be truly superconducting with
vanishing dc resistance.
A popular recent theory predicts for weak disorder 
a first order transition to a  Bragg glass state,
which is characterized by slow, at most algebraic, decay of translational
crystalline order \cite{giamarchi&doussal}. 

With this inconclusive theoretical background, experimental and
numerical evidence have had a major impact on the picture of the 
phase diagram of high temperature superconductors (HTSC)  
including fluctuations and disorder. A popular  
phase diagram including disorder which accounts
for many experimental features, notably the
loss of first order behavior at high fields, is shown in
Fig.\ \ref{fig:compd}(b). Our numerical results 
with disorder give a phase diagram with only one phase, 
a vortex liquid, just as in the clean limit.

\subsection{Experimental evidence}
\label{sec:ybcoexp}

This section  attempts a review of experimental 
evidence on which both our and more conventional pictures of the phase 
diagram of layered superconductors are based.
We discuss only  evidence in YBCO, because 
this is the material to
which our numerical model applies naturally.

\subsubsection{First order transition}
There is striking experimental evidence for a first order transition
in YBCO. The earlier evidence for discontinuous behavior
suggesting a first order transition came from resistive measurements
\cite{Kwok-dyn}. A sharp drop in resistivity 
was found to occur  at a temperature well below the $H_{c2}$ line.
Later it was shown that these resistive drops coincide with a 
discontinuity in the magnetization, the first
thermodynamic quantity found to be discontinuous at 
the transition line \cite{welp_melt-dyn,Liang}. 
The first measurements  of the latent heat which unambiguously
characterises a first order transition, 
were made by Schilling {\it et al.}\ \cite{schilling_melt} in  
1996. Since then a latent heat at the 
first order vortex transition has been  observed 
in different crystals of YBa$_{2}$Cu$_{3}$O$_{7-\delta}$ 
with varying oxygen deficiencies $\delta$ 
\cite{Junod_melt,Roulin_endpt} and for different orientations 
of the applied field \cite{schilling_anisotr}.  
The $B(T)$ dependence of the first order transition line obeys 
the standard continuum anisotropic scaling rules 
\cite{blatter-anisotr}  under 
rotation of the applied field away from the $c$-axis 
\cite{schilling_anisotr}. 

The scaling behavior of the first order transition lines for 
samples with different
oxygen deficiencies $\delta$ and therefore  different mass 
anisotropies $\gamma$ is of some interest as it can 
be easily compared to predictions of different theoretical models.
In a range of samples the first order lines have been found
to scale with $1/\gamma$ by Roulin {\it et al.}\ \cite{Roulin_endpt}, 
i.e.\ $\gamma  B(T)$ collapses on
one scaling curve. This is in disagreement with standard 
London-Lindemann type vortex lattice melting theory, which  
predicts the melting curve to scale inversly with the Ginsburg number
$Gi$ as
$1/Gi \propto 1/\gamma^2$ \cite{blatterrev}. The $1/\gamma$ 
scaling form is consistent with 3D lowest Landau level (LLL) 
scaling and our numerical results.

\subsubsection{Loss of first order behavior}
The first order behavior at the vortex transition 
has been observed to vanish at an upper critical 
field $B_{uc}$ for samples which are 
not fully oxygenated \cite{Roulin_endpt}. $B_{uc}$ is found to  
increase with decreasing concentration of oxygen deficiencies $\delta$
in YBa$_{2}$Cu$_{3}$O$_{7-\delta}$.
For $\delta\!=\!0$ a latent heat can be observed up to the 
highest experimentally investigated fields of $16$T 
\cite{Junod_melt}.  The end of the first order line
thus appears to be strongly correlated with the amount of
point disorder in the form of oxygen vacancies present in the 
system. The upwards shift of $B_{uc}$ with increasing 
oxygen content 
fits in well with the theoretical phase diagram 
in Fig.\ \ref{fig:compd}(b), where it corresponds to an extension of the Bragg 
glass phase to higher fields with decreasing disorder. 
 
Another striking feature of the experimental first order transition 
line in YBCO is its termination at low magnetic fields, which 
has been consistently observed 
in all relevant calorimetric measurements 
\cite{schilling_melt,Junod_melt,Roulin_endpt,schilling_anisotr}. 
The latent heat disappears for  
fields smaller than some lower critical field $B_{lc}$.  
The existence of a low field end-point is usually found
``puzzling'' \cite{Roulin_endpt}. 
The variations of  $B_{lc}$ for
specific heat measurements in different samples are large and 
qualitatively unexplained in the framework of a vortex lattice 
melting picture. 
For $B\!\parallel \!c$ in near optimally 
doped
samples  with a high level of oxygen deficiency  
$\delta > 0.06$, $B_{lc}$ is approximately  $0.7$T \cite{schilling_melt}. 
Measurements of $B_{lc}$ in different samples show that 
$B_{lc}$ increases with oxygen content \cite{Roulin_endpt},
which suggests at first sight a correlation with twin density. 
The value of $B_{lc}$ in fully oxygenated, twinned 
samples is of the order of several Tesla \cite{Junod_melt}.
The authors of Ref.\ \cite{Roulin_endpt} discuss the origin of the 
end-point and the variation of $B_{lc}$ in different samples.
The fact that detwinning does not noticeably change 
$B_{lc}$ and the existence of $B_{lc}$ in naturally untwinned samples, 
together with the reproducibility of $B_{lc}$ in different fully 
oxygenated samples, leads them
to the conclusion that an intrinsic mechanism  as a cause for the end
point cannot be excluded. 

An important relation for our discussion of the value of
$B_{lc}$ in different YBCO samples (see Sec.\ \ref{sec:critpt}) 
is that an increase in oxygen content corresponds
not only to an increase in natural twin density,  
but also to a systematic decrease in the anisotropy 
$\gamma$ in the samples used in Ref.\ 
\cite{Roulin_endpt,Roulin_vgmelt}. Our work
suggests that this change in anisotropy
rather than the presence of twins may cause the change in 
$B_{lc}$. 
Our numerical work provides an explanation for the existence
of $B_{lc}$ as well as a qualitatively correct prediction of its
rapid increase when $\gamma$ is decreased, such as can be
achieved by increasing the oxygen content.
Another noteworthy point which we shall discuss in Sec.\ 
\ref{sec:irrev} is that the location of $B_{lc}$ according to 
magnetization measurements is not always in agreement with the one  
measured in specific heat measurements.
In a fully oxygenated sample in Ref.\ \cite{Junod_melt} 
the latent heat vanishes at ca.\ 6T
while a magnetization discontinuity is still observed down 
to a field of 4T. 

Transport measurements reflect the loss of first order character of
the transition for low as well as for high fields 
\cite{Kwok-dyn,crabtree-rev}. The resistance 
only drops to zero, which would be the expected 
resistance for a weakly pinned
lattice at the very low voltages used, 
for a limited range of magnetic fields. For high and
low fields only a fractional drop is visible, which disappears
completely somewhat below 2 and above 7 Tesla. 

Below $B_{lc}$ and above $B_{uc}$ as well as in samples where no
latent heat at all is observed, a ``step'' in the heat capacity $C$ remains
\cite{schilling_melt,Junod_melt,Roulin_endpt,Roulin_vgmelt}.
This behavior has been interpreted  as evidence
for a second order transition. The sharpness of this 
``step'', is 
not altogether convincing 
(see e.g. Fig.\ 3 in Ref.\ \cite{schilling_anisotr}). 
However, the existence of a continuous 
transition to a vortex glass state at high fields 
is expected for the theoretical phase diagram in Fig.\ \ref{fig:compd}(b). 
According to the same  phase diagram
another line marking a field driven phase transition   
line from Bragg glass to vortex glass is expected to 
emerge where first order melting turns 
continuous \cite{giamarchi&doussal}. The  
``fishtail'' magnetization anomaly \cite{fishtail}, which correlates with the
location of $B_{uc}$ \cite{newfeat} could be interpreted  as evidence 
for such a transition.  A lack of sharpness of this feature 
makes it a candidate, however,  for a crossover rather than a phase transition.
There has also been evidence from resistive measurements
for a field driven crossover line as an extension of 
the first order transition in YBCO \cite{safar}.

\subsubsection{Evidence for a vortex lattice}

A vital ingredient of the vortex lattice melting
scenario which this paper disputes is the existence of a vortex 
lattice. Experimentally a vortex lattice is indistinguishable
from a liquid or glassy phase with short-range 
crystalline order on length scales large
compared to the vortex separation. 
Evidence for hexagonal 
coordination over large distances can be seen in YBCO for low fields 
in Bitter pattern decoration experiments 
\cite{gammel-bitpat}. At high fields this technique fails
because the vortices are too close to be individually resolved.
A powerful method to detect long-range vortex correlations
is neutron scattering \cite{bham-neutscat}. The Bragg peaks
observed in these experiments show that vortex positions 
are long-range correlated in all directions. 
The correlation length along the field can be enhanced 
by twin boundaries if the
field is oriented along the $c$-axis. However, data
from experiments with different orientation of the 
applied magnetic field show similar results, 
and thus indicate that the long correlation lengths 
along the field are independent of the presence of twin planes. 
The intrinsic crystalline in-plane correlation length 
is more difficult to deduce from neutron scattering data, because 
twin boundaries and/or pinning to the underlying crystal 
determine preferred orientations and can thereby strongly
enhance orientational order \cite{bham-anisot}. 

Although neutron scattering experiments give evidence
for long-range vortex correlations, some features of the data are 
unexpected in the framework of a 
vortex lattice melting picture. The observed diffraction patterns
suggest the existence of a vortex lattice or a Bragg 
glass, which means that the melting transition is expected to 
be of first order.  Such a first order melting transition 
should be visible as a discontinuous appearance of
Bragg peak intensity, as the temperature is reduced. 
However, the peaks appear continuously, 
which leads the authors of Ref.\ \cite{bham-neutscat}
to the conclusion that they must be dealing with second order 
vortex glass melting. 

An additional feature of the melting line defined by the onset of 
Bragg peaks (which is not mentioned in Ref.\
\cite{bham-neutscat}) is that it lies in the $B$-$T$ phase
diagram distinctly below the  line 
corresponding to thermodynamically measured
first order transition lines \cite{schilling_melt,Roulin_endpt}
under the assumption of scaling with mass anisotropy like
$B\propto 1/\gamma$ or $B \propto 1/\gamma^2$.
This point will be investigated in more detail 
in Sec.\ \ref{sec:bragg}.
For the alternative phase diagram presented in this paper, 
a continuous onset of Bragg peak intensity 
somewhat below the first order transition line 
is just the expected behavior.

\subsection{Numerical simulations}

A large number of numerical Monte Carlo or Langevin dynamics 
simulations of different three dimensional models 
have with few exceptions 
provided evidence of first order vortex lattice melting. 
The main disagreement between different simulations 
of layered models concerns the question whether 
layer decoupling coincides with this melting transition.
In this section we give an overview of different
numerical results and point out what we consider to be 
the weaknesses of the respective models used.

From the frustrated XY and Villain and lattice London models 
there is evidence for first order melting and distinct decoupling
\cite{chen&teitel-ryu&stroud-hagenaars&brandt} as well as, at least 
in the thermodynamic limit, only one simultaneous 
first order melting and decoupling transition
\cite{nguyen&sudbo-koshelev-hu}.  
Very recent simulations of the uniformly frustrated 3D XY model 
show a vortex lattice melting transition as well as a second,
possibly first order, 
phase transition within the liquid phase, at which 
the vortex line tension goes to zero \cite{nguyen_new}. 
In these models vortices are confined to a lattice. 
In 3D, a lattice acts as a close grid of columnar pins with infinite
pinning potential in the thermodynamic limit 
which may lead to spurious phase transitions. 
For the LLL, alteration of the phase diagram by 
the presence of a lattice pinning potential which
breaks translational and rotational invariance has been 
predicted from a theoretical analysis \cite{ktitorov&al}.

There are also numerical models that avoid using a lattice.
Numerical models relying on the 2D Bose gas analogy 
yield simultaneous melting and loss of phase coherence
along the $c$-axis \cite{nordborg}.
A different scenario has  been seen in a simulation
by Wilkin and Jensen \cite{wilkin&jensen}, in
which vortex pancakes in different layers are represented 
by particles with in-layer short-range repulsive and
inter-layer attractive interactions. A first order transition associated
with decoupling of vortices and without melting character
is observed. At a lower temperature a melting crossover
without noticeable thermodynamic signature occurs.
When sufficient point disorder is added, the decoupling
transition loses its first order character.
While not being affected by pinning to a numerical
lattice, the latter models may give unrealistic
results because they  allow variation of vortex position only, 
neglecting fluctuations of order parameter magnitude and
in many cases 
having unrealistic short-range interactions \cite{blum-screening}. 
Simulations using  the Lawrence-Doniach (LD) model in the 
LLL limit, which allows for these fluctuations 
and which has long-range vortex interactions,  
show a single first order simultaneous melting and decoupling 
transition \cite{sas&str3D,hu&mcd3D}. 
All of the simulations mentioned in this paragraph use
periodic  boundary conditions perpendicular to the field, 
which we believe can also lead to unphysical results
(see Sec.\ \ref{sec:nummod}). 

In the following sections we introduce our numerical model 
(Sec.\ \ref{sec:nummod}) and report 
results from our simulation. Comparisons to experimental data on 
YBCO are made in each section in the context of 
the relevant numerical results. Sec.\ \ref{sec:numpd} addresses the
numerical phase diagram of layered systems in the clean 
limit, followed by an analysis of order parameter correlations 
in space and time in Sec.\ \ref{sec:opc}.  
Numerical results in the presence of quenched random disorder
are reported in Sec.\  \ref{sec:do3d}. 
The paper closes with a discussion and a summary of our work.

\section{Numerical Model}
\label{sec:nummod}

Our simulation of a layered superconductor uses the
Lawernce-Doniach (LD) model \cite{LD}, which 
consists of a stack of planes with Josephson coupling 
between neighboring layers.
With the superconducting order parameter in the $n^{th}$ layer denoted as 
$\psi_{n}$, the Hamiltonian for the layered system 
in a magnetic field perpendicular to the layers is
\begin{eqnarray}
\label{eq:ham}
&{\cal H}_{clean}&=\sum_{\hbox{n}}d_{0}\int\! d^{2}\!r 
        \left(\alpha|\psi_{n}|^{2}+\frac{\beta_{2D}}{2}|\psi_{n}|^{4}+
        \right.\nonumber\\
        && \left.\frac{1}{2m_{ab}}
               |(-i\hbar{\bf\nabla}\!-\!2e{\bf A})\psi_{n}|^2
              +J|\psi_{n+1}\!-\!\psi_{n}|^{2}
            \right),\nonumber
\end{eqnarray}
where $d$ is the layer periodicity, $d_0$ is the layer thickness and 
${\bf B}={\bf \nabla \times A}$, which we shall take as constant and
uniform.
The same Hamiltonian can be read as the finite difference
approximation to an anisotropic 
continuum model with $\psi(nd)=\sqrt{d/d_{0}}\psi_n$, 
$\beta=\beta_{2D}\times d/d_{0}$ and $m_c=\hbar^{2}/2 J d^{2}$.
In first approximation $\alpha(T)=\alpha '(T-T_{c})$
and $\beta_{2D}(T)$ is  constant;  $ \alpha ',\beta_{2D}, J>0$.

We simulate the LD model  with $N_{ab}$  vortices per layer 
in $N_{c}$ layers.
Along the $c$-axis  we use periodic boundary conditions.
In the $ab$-planes we chose a different, more unusual
approximation to the thermodynamic limit of an infinite plane.
The layers are taken to be of spherical geometry with a 
radial magnetic field. 
The reasons for our preference of this geometry to the more
widely used geometry of a plane with periodic boundary conditions 
have been discussed in detail by Dodgson and Moore \cite{dodgson}. 
The main advantage is that the
spherical geometry guarantees full rotational and translational 
symmetry, which periodic boundary conditions do not. 
One example where the spherical geometry captures the physics 
better than periodic boundary conditions --
despite the topological defects imposed on the triangular
lattice ground state on the sphere \cite{dodgson} --
are particles interacting with  the $1/r^{12}$ 
interaction. Here simulations on 
a sphere show already for moderate system sizes 
the genuine continuous transition
to the crystalline state \cite{perrez}, 
while with periodic boundary conditions a spurious first 
order transition occurs even for very large system sizes. 

For each layer $\psi$ is expanded in eigenstates of the squared momentum
operator $(-i\hbar {\bf \nabla}-2e{\bf A})^{2}$. We retain eigenstates
belonging only to the
lowest eigenvalue  $2eB\hbar$, (the LLL approximation)
which  is a useful procedure over a large portion of the vortex liquid 
regime \cite{ikeda}.
Our numerical model is an extension to the model of a spherical thin
film used in Ref.\ \cite{dodgson,kienappel,o'neill&lee}.
The magnetic potential is ${\bf A}=BR \;\tan(\theta/2) \hat{\phi}$,
for which an orthonormal basis of the LLL eigenfunctions in each layer
is given by 
\begin{equation}
 \phi_{m}=
\frac{A_{m}}{(4\pi R^{2})^{1/2}}
e^{im\phi}\sin^{m}(\theta /2)\cos^{N_{ab}-m}(\theta /2), 
\label{eq:lll}
\end{equation}
where $A_{m}=((N_{ab}+1)!/(N_{ab}-m)!m!)^{1/2}$
for $0\leq m\leq N_{ab}$.
Note that we use units of length $l_{m}=(\hbar/2eB)^{1/2}$, which 
fixes the sphere radius as $R=(N_{ab}/2)^{1/2}$. 

The order parameter in every layer is expanded in
the above basis set as  
\begin{equation}
\label{eq:phiexp}
{\psi_n(\theta,\phi)}=Q\sum_{m=0}^{N}v_{n,m}\phi_{m}(\theta,\phi),
\end{equation}
where $Q=(2 \pi k_{B}T/ \beta_{2D} d_{0})^{1/4}$.
Orthonormality of the LLL eigenfunctions 
can now be used to express the Hamiltonian in terms of 
the LLL coefficients $v_{n,m}$ and only two parameters,  
$\alpha_{2T}$ and $\eta$: 
\begin{eqnarray}
\label{eq:simham2}
\frac{{\cal H}_{clean}}{k_{B}T}
&=&\sum_{n=1}^{N_{c}}\left (\alpha_{2T}\sum_{m=0}^{N_{ab}}|v_{n,m}|^{2}
+\frac{1}{2N_{ab}}\sum_{p=0}^{2N_{ab}}|U_{n,p}|^{2}+
\right.\nonumber\\&&\left. 
|\alpha_{2T}\eta|  \sum_{m=0}^{N_{ab}} |v_{n+1,m}-v_{n,m}|^{2} \right),
\end{eqnarray}
with $U_{n,p}(\{v_{n,m}\})$ as defined in Sec.\ \ref{sec:quarten}.  
The 2D effective temperature and field parameter for each layer
is  $\alpha_{2T}=(d_{0}h/2e\, \beta_{2D}  B\, k_{B}T)^{1/2}\alpha_{H}$,
with $\alpha_{H}=\alpha(T)+eB\hbar/m_{ab}$.
The scaling parameter $\eta$ relates to the Josephson coupling 
constant $J$ as $\eta=J/|\alpha_{H}|$. 
We can define an effective mass $m_{c}$ via
$\eta=\hbar^2/2m_c d^2|\alpha_{H}|$. Then $\sqrt{\eta}$
is the ratio of the 3D mean-field coherence length to the
layer periodicity,  $\sqrt{\eta}=\xi_{\parallel}/d$.
Note that for a HTSC material the 2D parameters 
$\beta_{2D}$ and $d_{0}$ are effective microscopic properties
of the copper oxide layers and  essentially unknown.
However, they enter the simulation only via 
$\alpha_{2T}=(2\pi d_{0}/ \beta_{2D} \, k_{B}T)^{1/2}\alpha_{H}$
where they can be replaced by the layer periodicity $d$ and 
the bulk $\beta$ using the relation 
$\beta=d/d_{0} \times \beta_{2D}$.                          

The state in the LLL-LD model depends on 
two dimensionless scaling parameters,
$\alpha_{2T}$ and $\eta$. It is useful to have 
two scaling parameters which can be thought of as 
something physical, e.g.\ one characterising 
temperature and the other coupling strength between layers.  In this sense, 
$\alpha_{2T}$ and $\eta$ are not very appropriate. 
Because $d_{0}<d$, the temperature parameter $\alpha_{2T}$ 
goes to zero independently  of $B$ and $T$ in the $d\rightarrow 0$ 
limit of a continuous system. 
The coupling parameter $\eta$  includes 
a factor $|1/\alpha_{H}|$,  which means that it diverges at the 
mean-field $H_{c2}$. For these reasons
we choose as effective temperature and coupling strength
two different parameters  that depend on  
$\alpha_{2T}$ and $\eta$. 

For the temperature parameter in a layered model describing a 
bulk sample, the 3D version of the LLL scaling variable $\alpha_{T}$ 
\cite{thouless&ruggeri} stands out as an  
appropriate candidate. It is given by 
\begin{equation}
\label{eq:defalphat}
\alpha_{T}\!=\!\left(\frac{\sqrt{2}\hbar^{3/2}\pi}
             {k_{B} e^{3/2}\mu_{0}}\right)^{2/3}\!\! 
             \left( \frac{1}{\kappa^{2}\gamma}\right)^{2/3}
            \frac{\partial B_{c2}/\partial T |_{T_{c}}(T-T_{c})+B}
                                 {(BT)^{2/3}},
\end{equation}
which can be expressed in terms 
of $\alpha_{2T}$ and $\eta$ as 
$|\alpha_{T}|^{3}= \eta (2\alpha_{2T})^{4}$. At low temperatures 
$|\alpha_{T}|^{3/2}$ behaves as 1/T.
A good measure of
coupling strength is the product $|\alpha_{2T}\eta|$ which multiplies
the coupling term in the Hamiltonian. It is in SI units
given by 
\begin{equation}
\label{eq:deflc}
|\alpha_{2T}\eta|=
  \left(\frac{\hbar^{3} \pi}{8 e^{3} k_{B} \mu_{0}}\right)^{1/2}
        \frac{1}{\kappa \gamma^{2} d^{3/2}}
        \frac{1}{(BT)^{1/2}}. 
\end{equation}
Other than the  factor $1/\sqrt{BT}$, $|\alpha_{2T}\eta|$ contains 
only constants and therefore varies slowly for rather 
a wide range of $\alpha_{H}$. It can thus be regarded
as a material constant over considerable regions of the $B$-$T$ 
plane. The limiting case $|\alpha_{2T}\eta|\rightarrow 0$ 
describes the 2D system, while small $|\alpha_{2T}\eta|$ means strongly
layered characteristics. For constant $\alpha_{T}$,
the limit $|\alpha_{2T}\eta| \rightarrow \infty$  is the continuum limit,
in which all system properties depend on $\alpha_{T}$ alone.
Note that our model parameters depend on the bulk material 
parameters $\kappa$,  mass anisotropy $\gamma$,  layer separation $d$,
$\partial B_{c2}/\partial T |_{T=T_{c}}$,  $T_{c}$
as well as $B$ and $T$ only. All
of these are for HTSC more or less well known from
experiment.

Most of the experimental evidence to which we can compare
our results is from samples which have at least weak 
disorder due to crystal defects 
or impurities. An idealized version of this kind of disorder 
is a random local variation of $T_{c}$.  
To simulate the effects of such quenched random disorder, 
a random local potential $\Theta$ can be  added to the quadratic energy 
term of every layer, to give a disorder contribution to the
Hamiltonian of   
\begin{equation}
\label{eq:dis_fe}
{\cal H}_{dis}
=d_{0}\int d^{2}r \,\Theta({\bf r})|\psi({\bf r})|^2,
\end {equation}
where $\Theta({\bf r})$ is real and  Gaussian distributed with 
\[
\overline{\Theta({\bf r})}=0,
\]
\begin{equation}
\label{eq:deltadef}
\overline{\Theta({\bf r})\Theta({\bf r'})}=
               \Delta\delta({\bf r}-{\bf r'}).
\end{equation}
Here $\delta$ is the two dimensional Dirac delta 
function and $\Delta$ is the measure of the strength of the disorder.
The random realizations of $\Theta$ are different  
in every layer, and ${\cal H}_{dis}$ is the sum of the single layer 
contributions. The disorder contribution ${\cal H}_{dis}$  
in terms of the LLL coefficients as used in our simulation 
is given in the Appendix, Sec.\ \ref{sec:append2}.
Unfortunately there is no experimental measure of $\Delta$, and we can 
therefore not express the strength of disorder in terms of 
measurable quantities.

As in reference \cite{kienappel},  our simulation follows 
Langevin (model A) dynamics. We drive our system by the time 
dependent Ginsburg-Landau equation, 
discretized in time and expanded in the appropriate eigenfunctions:
\begin{equation}
\label{eq:tdev}
v_{n,m}(t\!+\!\Delta t)-v_{n,m}(t) 
        =-\Delta t\,\Gamma \frac{\partial{\cal H}(t)}
          {\partial v_{n,m}^{\ast}}
        +\Delta t\,\xi_{n,m}(t).
\end{equation}
The complex random noise variables $\xi_{n,m}$ are drawn independently from 
a Gaussian probability distribution, so that their magnitude 
has a  variance $\sigma^2/\Delta t=1$, where
$\sigma^2=2\,\Gamma\,k_{B}T$, so that $\Delta t$
is the only free parameter. We chose $\Delta t=0.15$ 
(see Ref.\ \cite{kienappel}).

\section{Numerical Phase Diagram}
\label{sec:numpd}
\vspace{-0.5cm}
\begin{figure}
\centerline{\epsfxsize= 7 cm\epsfbox{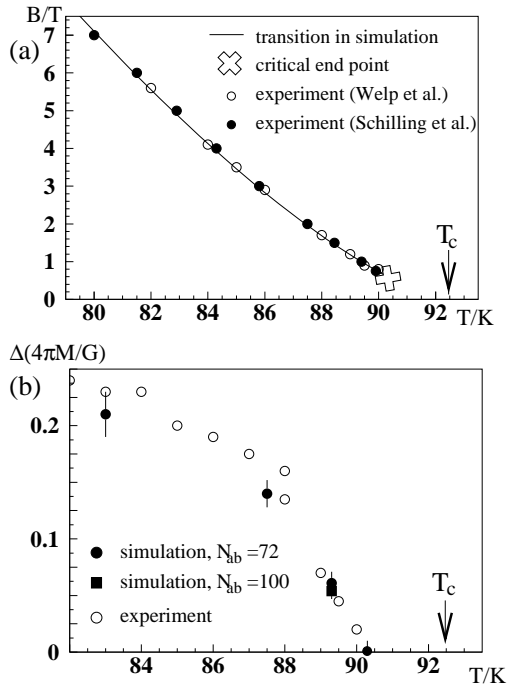}}
  \caption{Phase diagram (a)  and magnetization discontinuity 
   (b). Experimental 
   data is taken from Ref.\ \protect\cite{welp_melt-dyn,schilling_melt} (a)
   and \protect\cite{schilling_melt} (b). (For $N_c$ see section 
   \protect \ref{sec:simfot})}
  \label{fig:btdiamag}
\end{figure}
As an introduction to our numerical results we show in 
Fig.\ \ref{fig:btdiamag}(a) our numerical phase diagram 
for the clean case in comparison to experimental results. The
only phase present in this phase diagram is a vortex liquid. 
We see a first order transition
line between two vortex liquid states 
with  a critical end-point at low fields, which 
agrees well with the experimental YBCO 
``melting'' line. 
The magnetization jumps we observe are 
shown in Fig.\ \ref{fig:btdiamag}(b). They are in very good agreement with
data for YBCO from Ref.\ \cite{schilling_melt}.  The 
magnetization jumps in these experimental measurements are very 
likely to be of thermodynamic origin, as they are found to 
be consistent (according to the
Clausius Clapeyron relation)
with the latent heat data measured in the same sample 
\cite{schilling_melt}.
Figure \ref{fig:btdiamag}(a) and (b) 
were obtained using standard 
YBCO values for the fitting parameters;
viz for the Landau-Ginsburg 
parameter $\kappa$=60, the mass anisotropy $\gamma$=7.5, 
the slope of the mean-field transition line
$\partial B_{c2}/\partial T |_{T=T_{c}}=-$2T/K, 
the mean-field $T_{c}$=92.5K and the layer separation $d$=11.4\AA.

\subsection{Continuum limit}

\begin{figure}
\centerline{\epsfxsize=9cm\epsfbox{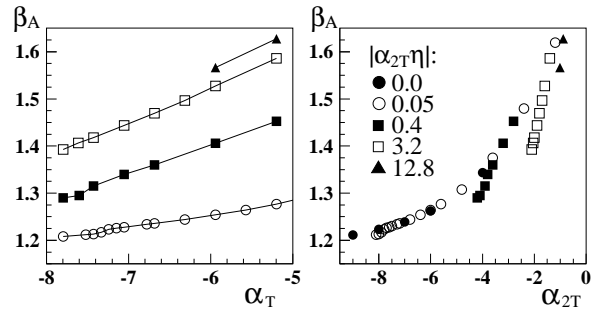}}
  \caption{Abrikosov ratio $\beta_{A}$ for a range of 
  coupling strengths $|\alpha_{2T}\eta|$ plotted against the 3D and 2D
  effective temperature variables $\alpha_{T}$ and $\alpha_{2T}$.
  Solid lines are guide to the eye.} 
  \label{fig:ba}
\end{figure}

From LLL scaling \cite{thouless&ruggeri} we know that all thermodynamic
properties depend on  $\alpha_{2T}$ alone in the 2D limit 
($|\alpha_{2T}\eta|\rightarrow 0$) and on $\alpha_{T}$
alone in the continuum limit ($|\alpha_{2T}\eta|\rightarrow \infty$).
Figure \ref{fig:ba} shows how the thermal average of a typical
quantity of interest, here the Abrikosov ratio
$\beta_{A}=\langle|\psi|^4\rangle/\langle|\psi|^2\rangle^2$,
behaves in the intermediate regime of a positive finite coupling
strength.
In the high temperature regime 
2D scaling applies, in the low temperature regime however 3D
scaling becomes more appropriate. If we look at our model 
as a finite difference approximation to the continuum case,
this means that this approximation is for the same 
layer spacing  $d/\xi_{||}=1/\sqrt{\eta}$ better at low 
than at high 3D temperatures $\alpha_{T}$.
This is a natural result as correlations along the field direction 
increase in the layered system as $\alpha_{T}$ decreases. 
We are, however, not
able to simulate system sizes that behave fully continuum-like at 
moderate temperatures. The numbers of layers used for the data 
in Fig.\ \ref{fig:ba} are chosen to have the correlations along 
the $c$-axis distinctly 
smaller than the system size, which means for $|\alpha_{2T}\eta|=12.8$
even at the moderate temperature of $\alpha_{T}\!=\!-6$ 
an $N_{c}$ of 200.

\subsection{First order transition}
\label{sec:simfot}

\begin{figure}
\centerline{\epsfxsize=9cm\epsfbox{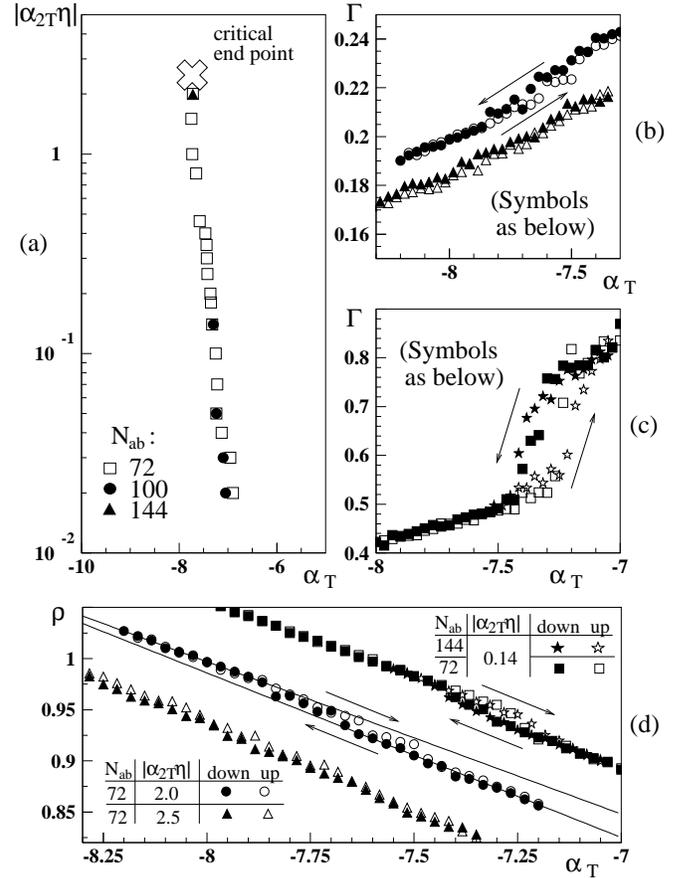}}
  \caption{Plot (a) shows
  the numerical phase diagram. First order transition points 
 are plotted in the $\alpha_{T}\!-\!|\alpha_{2T}\eta|$ plane. 
  Plots (b)-(d) show order parameter density 
  $\rho$ and the degree of layer independence $\Gamma$ upon
   heating and cooling. Note the hysteresis in the system, the 
   clear first order behavior for $|\alpha_{2T}\eta|=2$ 
   (solid lines are linear fits of $\rho$ above and below the
   transition)  and the lack of first order behavior for 
   $|\alpha_{2T}\eta|=2.5$. $N_{c}$ varies between 8 and 80 for
   $|\alpha_{2T}\eta|$ between 0.02 and 2.5. For $|\alpha_{2T}\eta|=2.5$,
   $\rho$ is offset by $-0.05$.}
\label{fig:alph}
\end{figure}

Figure \ref{fig:alph}(a) shows the phase diagram 
in terms of simulation parameters, where data points mark the
location of first order transition points as found upon heating and cooling
the system. The logarithmic scale is chosen for even data distribution.
As  $|\alpha_{2T}\eta|$ increases
and the system approaches the continuum limit, the transition line 
terminates at a critical end-point. 
Note that along the transition line $\alpha_{T}$ is approximately 
constant, which means that the
field and temperature dependence of the transition line behave as 
expected for a
continuum model where $\alpha_{T}$ is the only scaling parameter in
the system. 
Another point of interest is  the limit 
$|\alpha_{2T}\eta| \rightarrow 0$ of independent 2D layers.
If we approximate the transition value of $\alpha_{T}$ as 
constant for different $|\alpha_{2T}\eta|$, this translates 
to a dependence of $\alpha_{2T}$ on $|\alpha_{2T}\eta|$ 
along the transition line as 
$\alpha_{2T}\propto -|\alpha_{2T}\eta|^{-1/3}$. 
As the coupling is reduced to zero, the 2D transition temperature
diverges, $\alpha_{2T} \rightarrow -\infty$.
This would imply that there is no finite temperature phase 
transition in two dimensions, in agreement with 2D simulation 
results   
\cite{dodgson,o'neill&lee,kienappel}. Such behavior would be a natural for  
a transition of a predominantly decoupling nature, which cannot 
occur in thin films. However, we find that direct extension of our 
numerical transition line to even lower couplings and thereby 
numerical observation of the 2D limiting behavior
is impossible due to the finite size effects analyzed in 
Appendix \ref{sec:numlim}.

Figures \ref{fig:alph}(b)-(d) show examples of the kind 
of measurement used to locate the first order transition.
The system displays hysteresis upon heating and cooling. 
We mark the transition in the middle of the observed hysteresis 
loop. The hysteresis decreases with sweeping rate (typically 
10,000-20,000 time steps per data point), and for a few 
cases we have confirmed with equilibrium measurements that the 
equilibrium transition 
coincides roughly with the middle of the hysteresis loop.
The coupling values for which we show hysteresis measurements
correspond in decreasing order firstly to the critical end-point, 
secondly the nearest to the critical end-point where we have measured 
hysteresis, and thirdly an arbitrary, low coupling value. 
The discontinuity $\Delta\rho$ in the order parameter 
density $\rho$, given by 
$\rho=(\alpha_{T}\, \beta/2 \pi \alpha_{H}) \times \langle |\psi|^{2}\rangle$,
is found to be more or less constant between 
$|\alpha_{2T}\eta|=$0.1 and 1.5. A decrease in $\Delta\rho$ 
is observed below $|\alpha_{2T}\eta|=$0.1. However, this decrease at
low couplings is 
possibly due to finite size effects, which are in detail 
described in the Appendix
\ref{sec:numlim}. The rapid decrease to zero between  
$|\alpha_{2T}\eta|=$1.5 and $|\alpha_{2T}\eta|=$2.5
appears system size independent.

The plots of the hysteresis in the degree of independence of 
neighboring  layers, given by 
\[ \Gamma=\frac{\langle |\psi({\bf r})-\psi({\bf r}+d\hat{c})|^2\rangle}
              {\langle |\psi|^2\rangle} \; ,
\]
reveal more about the nature of the transition and its disappearing.
For low coupling, there is a large jump in $\Gamma$ at the
transition, which decreases throughout parameter space
until it is very small just before the end-point. 
The decoupling character of the transition gradually
decreases along the transition line until it disappears at 
the critical end-point. Further discussion predicting the existence and
approximate location of the critical end-point from the decoupling 
character of the transition and more detailed  numerical results 
concerning the critical end-point will be given in Sec.
\ref{sec:critpt}. The reader should note that in our simulation the  
onset of decoupling does not imply the onset of a superfluid 
density at the transition.

The magnetization discontinuities in Fig.\ \ref{fig:btdiamag}(b)
are calculated from the discontinuities in $\rho$.
The magnetization in the LLL model is  
$4\pi M=-(\mu_{0}e\hbar/m_{ab})\langle |\psi|^{2}\rangle$, 
which is in terms
of our simulation parameters 
$4\pi M=\pi(B-B_{c2}(T)) \rho/\alpha_{T}\kappa^{2}$,
where $B$ is the applied magnetic field.
Thus we can work out the magnetization discontinuity from the
discontinuity in $\rho$ taken from the two linear extrapolations  
at the transition.
The data points in Fig.\ \ref{fig:btdiamag}(b) represent  
$|\alpha_{2T}\eta|=$1, 1.5, 2 and 2.5 for $N_{c}=$50, 60, 60/80, 80.
For these four transition  points we have an average value of 
$\alpha_{T}= -7.72$, which yields the transition line in 
Fig.\ \ref{fig:btdiamag}.

The magnetization discontinuity is thermodynamically linked to the 
entropy change at the transition by the Clausius-Clapeyron relation
\begin{equation}
\label{eq:clclp}
\Delta S=-\Delta M \frac{dH_{FOT}}{dT}\; ,
\end{equation}
where $dH_{FOT}/dT$ is the slope of the first order 
transition line.
We cannot measure $\Delta S$ directly. However, 
the good agreement in   $\Delta M$ and the first order transition line 
between simulation and experiment on the one hand 
and the consistency of the experimental 
$\Delta S$  as calculated from magnetization and latent  
measurements for the samples in Ref.\  
\cite{schilling_melt} on the other hand 
imply agreement for $\Delta S$ between
simulation and experiment.
Note also the clear change in slope of the linear fits 
to $\rho$ at the transition. Locally $\alpha_T$
is linear in $T$, and via Maxwell's relation
\[
\left. \frac{\partial S}{\partial H} \right|_T=
\left.\frac{\partial M}{\partial T} \right|_H
\] 
the sudden change in slope of $\rho$ 
at the transition implies a change in slope of the 
entropy and a step-like feature
in the specific heat $C=T (\partial S/\partial T)_B$,
with a lower value on the low temperature side of the transition. 
This is consistent with experiment. The relative change in slope
we find from the fits in Fig.\ \ref{fig:alph} is 
8\%, while the equivalent experimental change in the 
heat capacity as taken from Fig.\ 3 of reference \cite{schilling_anisotr}
is of the order of 5\% (for a derivation of this value
see Sec.\ \ref{sec:beycp}). The fact that our simulation gives
evidence for a step in the specific heat is consistent 
with results from a theoretical 
analysis showing that the step seen in experiment can be 
accounted for by thermal fluctuations within the LLL 
approximation \cite{pierson}.

\subsection{The critical end-point}
\label{sec:critpt}
Until now we have as evidence for a critical end-point only the
fact that the hysteresis along the transition line 
eventually becomes unobservable.
To more firmly establish its existence, we shall consider
how the nature of the transition may lead to a critical end-point.

\subsubsection{Why does the first order transition disappear?}
A well known first order phase transition which ends at a 
critical end-point is that of the ordinary liquid-gas
transition. Here  the phase transition separates a liquid state with small
interparticle separation $d_{l}$ which takes advantage of the attractive
interparticle energy which exist at distances $d_{min}$, so
$d_{l}\approx d_{min}$, from a gaseous state
with large interparticle separations $d_g$ which is favored by  a 
high entropy. If the
density is increased to the point where it reduces $d_g$ to be of order
$d_{min}$, the transition line ends. We believe that in our case the entropy
advantage of the high  temperature phase arises
 when the order parameter values
in adjacent layers are uncorrelated i.e when the layers are decoupled.
The ratio of the mean-field coherence
length perpendicular to the layers to the  layer distance, 
$\xi_{||}/d=\sqrt{\eta}$, increases along the  transition line with increasing
coupling  parameter  $|\alpha_{2T}\eta|$. Because $\xi_{||}$ defines the
minimal extent of  order parameter correlations, a high-entropy state
with decoupled 
layers is not possible if $\xi_{||}\geq d$. And indeed, we 
will in the next section locate the critical end-point where the transition
disappears at  $|\alpha_{2T}\eta|\!=\!2.55$ and  $\alpha_T\!=\!-7.75$, 
which corresponds to $\sqrt{\eta}=1.06$.   
Very near the zero-field transition temperature 
$T_{c}$, where $\xi_{||} \gg d$, the system can be expected to 
behave like a continuum and thus a decoupling line cannot 
be expected to reach $T_{c}$.

\subsubsection{Divergence of length scales}

\begin{figure}
\centerline{\epsfxsize= 8cm\epsfbox{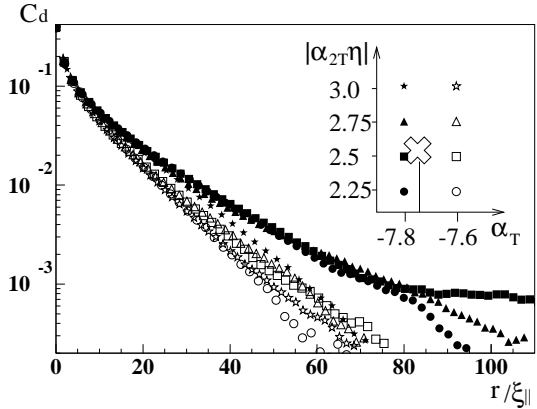}}
  \caption{Density-density correlations along the $c$-axis near the end 
           the first order transition line and the critical end-point (see inset). 
           Note the decreasing difference in correlations between 
           $\alpha_{T}=-7.6$ and $\alpha_{T}=-7.8$ as the transition
           disappears. For $|\alpha_{2T}\eta|=2.5$ and $\alpha_{T}=-7.8$
           we see evidence for a long length scale associated with
           fluctuations in the average order parameter density.
           System sizes are $N_{ab}$=72, $220<N_{c}<260$ for  
           $\alpha_{T}=-7.6$ and  $N_{c}=270$  for $\alpha_{T}=-7.8$.
        }
  \label{fig:zd}
\end{figure}

Near a critical end-point we do not only expect 
all discontinuities to disappear, but we also expect there to 
be a divergence of
the length scale of fluctuations in the order parameter
density of the system. 
We therefore looked at the density-density correlations 
of the order parameter, 
\begin{eqnarray}
\label{eq:dencordef}
\lefteqn{C_d({\bf r'},t')=}&& \\ &&
             \frac{\langle|\psi({\bf r},t)|^{2}
                       |\psi({\bf r}+{\bf r'},t+t')|^{2}\rangle
              -\langle|\psi({\bf r})|^{2}\rangle
               \langle|\psi({\bf r}+{\bf r'})|^{2}\rangle}
               {\langle|\psi|^{2}\rangle^{2}}\nonumber
\end{eqnarray}
in the case where ${\bf r'}$ is a vector parallel to  
the $c$-axis and $t'\!=\!0$. This correlator is expressed 
in terms of thermal averages of 
the LLL coefficients in the appendix, Sec.\ \ref{sec:3Ddc}.
Plots of these correlations near the critical end-point
can be seen in Fig.\ \ref{fig:zd}. 

There is evidence
of two length scales in the vicinity of the end-point.
The short distance decay of the  correlation function is 
dominated by the positional
correlations of the vortices in the different layers. This length scale
is mostly determined by $\alpha_{T}$ and changes slowly in the vicinity of  
the critical end-point. The diverging length scale 
is thus not that of positional correlations of the vortices,
but instead is associated with local density fluctuations.
This is not surprising if the analogy of an ordinary liquid-gas 
transition is considered, where crystalline correlations 
on a microscopic scale correspond to positional vortex correlations.
A noteworthy feature of the short distance decay
of vortex correlations is that the difference between 
curves at $\alpha_{T}=-$7.6 and $\alpha_{T}=-$7.8
decreases as the coupling parameter $|\alpha_{2T}\eta|$ 
is raised past the end-point, as one would expect for a 
disappearing discontinuity.

Density correlations diverge at the liquid-gas critical end-point 
on a mesoscopic scale. Analogous correlations on a larger scale
appear in the numerical data between $\alpha_{T}=-$7.6 
and $\alpha_{T}=-$7.8 as $|\alpha_{2T}\eta|$ is increased to its 
critical end-point value. Figure  \ref{fig:zd}
shows for $|\alpha_{2T}\eta|=$2.5 and $\alpha_{T}=-$7.8 evidence of 
a second, much longer length scale governing the decay of the 
correlation function at large distances. This length scale is associated 
with the density fluctuations at the critical end-point and  only becomes 
visible once it is larger than that of the vortex 
correlations. Due to the small amplitude of these density fluctuations 
very long simulation 
times are needed to see the correlations within the  
statistical noise.

\subsection{Anisotropic scaling and the value of $B_{lc}$}

The location of the numerical first order transition 
is to first approximation a line of constant  $\alpha_{T}$
and thus in agreement with 3D LLL scaling. 
This is, although surprising for very low couplings, 
not unnatural in YBCO, where even for the highest fields
and lowest couplings, e.g for 
a transition  at $B$=20T, $T$=65K,  $|\alpha_{T}\eta|$ is 
of the order $1/2$ and the correlation length along the $c$-axis 
above the transition (see Sec.\ \ref{sec:corrlen}) of the order of 
10 layers. 
The continuum approach can thus be expected to work fairly well
for YBCO. 

\subsubsection{Scaling with anisotropy $\gamma$}
According to Eq.\ \ref{eq:defalphat} the $B(T)$ dependence
of a line of constant  $\alpha_{T}$ ban be approximated as 
\[B \propto  \frac{(T_{c}-T)^{3/2}}{\kappa^{2}\gamma}.\]
Thus we expect our transition line in samples of different 
anisotropies to scale as $B \propto 1/\gamma$,
which is the experimentally observed scaling \cite{Roulin_endpt}.
This form of scaling disagrees with a London-Lindemann type melting 
theory. The variation of the location of the low field critical end-point
with anisotropy $\gamma$ is discussed at the end of this section in 
the context of a general analysis of variations in $B_{lc}$.

\subsubsection{Variation with the angle of the applied magnetic field}
We cannot in our simulation change the orientation of the 
applied field. We can, however, using the 
known properties of the numerical transition 
discuss the expected behavior upon such a change in field orientation. 
We believe that the first order transition is predominantly 
of a decoupling nature. However, it is important to keep in mind that 
the coupling between layers in our simulation is not magnetic coupling 
of vortex segments, but Josephson coupling of the order 
parameter. This type of coupling is independent of the orientation 
of the flux lines with respect to the $c$-axis. 
Under angle
rotation between $\theta\!=\!0$ ($B\!\parallel \!c$) and 
$\theta\!=\!\pi/2$ ($B\!\perp \!c$), the transition line 
should scale according to 
anisotropic continuum scaling \cite{blatter-anisotr} with $\theta$ as 
\[B(\theta,T)=\gamma_{\theta}B(0,T),\]
where $\gamma_{\theta}$ is given by 
$\gamma_{\theta}=(\cos^2\!\theta + \sin^2\!\theta\,/\,\gamma^2)^{-1/2}$.
This form of scaling has been observed for the first 
order transition line in YBCO \cite{schilling_anisotr}.

Our argument linking the nature of the transition  
to the location of the end-point applies for any orientation
of the magnetic field. 
The location of the end-point is for all orientations 
given by 
$\xi_{c}\approx d$, where $\xi_{c}$ is the coherence length along the 
$c$-axis. 
For $B\!\parallel \!c$  this condition is equivalent to 
$\xi_{||}\approx d$. Using
$\xi_{||}=\hbar/(2m_{c}|\alpha_{H}|)^{1/2}$, 
this condition can be transformed 
to a simple $B(T)$ dependence which scales
like the transition line itself. The end-point where both lines
cross therefore equally just shifts to a higher field as 
$B_{lc}(\theta)=\gamma_{\theta}B_{lc}(0)$. This form 
of scaling has been  experimentally observed 
for the end-point in YBCO \cite{schilling_anisotr}. 

\subsubsection{Variations in $B_{lc}$}

The location of the numerical critical end-point 
agrees well with the experimental data of Fig.\ \ref{fig:btdiamag}.  
However, as already mentioned in Sec.
\ref{sec:ybcoexp}, the experimental 
value of $B_{lc}$ varies widely 
between different samples. Approximating  $\kappa$ and $T$ 
as constant at the critical end-point, we obtain from Eq.\ 
\ref{eq:deflc} the scaling relation $B_{lc} \propto 1/\gamma^4$.
The $\kappa$ dependence
is $B_{lc} \propto 1/\kappa^2$.
These two scaling relations show that $B_{lc}$ depends
very sensitively on material parameters.

We can  use the fit in Fig.\ \ref{fig:btdiamag} with  
$\gamma=7.5$ and $B_{lc}=0.7$T as reference point to compare
the location of the critical end-point in our simulation to 
experimental values of $B_{lc}$. 
For samples with measured anisotropies of 
$\gamma$=7.8 (Ref.\ \cite{schilling_anisotr}), 
$\gamma$=7.0, 5.9, 5.3 (Ref.\ \cite{Roulin_vgmelt}) 
specific heat peaks have been measured 
down to  $B_{lc}$=0.7T (Ref.\ \cite{schilling_anisotr}),
$B_{lc}$=3T, 4.5T, 6T (Ref.\ \cite{Roulin_vgmelt}).
The rapid increase of $B_{lc}$ with decreasing anisotropy 
is in qualitative agreement with a scaling 
law $B_{lc}\propto 1/\gamma^4$.

Quantitative analysis however yields only poor agreement. 
The predictions using the above reference point and scaling law 
deviate from the experimental value by
$-15\%$ (Ref.\ \cite{schilling_anisotr})  
and  $-70\%$, $-60\%$, $-50\%$ (Ref.\ \cite{Roulin_vgmelt}).
There are many possible reasons other than variations in 
$\kappa$ for this quantitative
disagreement. Firstly, the finite width of the transition due to 
sample inhomogeneities may lead to a spreading out of the
specific heat peak, which can make it undetectable for the lowest fields
above $B_{lc}$ and lead to overestimates of $B_{lc}$.
Secondly any aspects of the physical coupling which are not 
represented in our model could lead to corrections in the 
effective $|\alpha_{2T}\eta|$ and should thus 
be included for an accurate description.

A third, very important point is that near $T_{c}$  critical
fluctuation effects  arising from the zero field 
transition are not negligible and especially 
affect the divergence of $\xi_{||}$. Such effects extend
to fields of the order of  
$Gi \times H_{c2}(0)$ \cite{tesanovicnew}, in YBCO $\sim$ 1T.
Up to these fields, the LLL approximation is invalid because 
higher Landau levels are needed to allow for critical
fluctuations. Thus the end-point lies in a region where 
the LLL approximation is inadequate, and we cannot 
expect our estimates of the  
position of the critical end-point to be quantitatively 
accurate. However, we can expect that such fluctuations
well make $B_{lc}$ smaller than estimated from the LLL approximation.
Critical fluctuations cause $\xi_{||}$ to diverge faster 
than in the LLL approximation, so it reaches a value $O(d)$
at temperatures closer to $T_{c}$ than in the LLL approximation.
This in turn will reduce the value of $B_{lc}$.

In addition, despite extensive finite size effect analysis
(see appendix \ref{sec:numlim}) we can never fully exclude the possibility
that the location of the end-point would shift to lower fields if we 
used much larger systems. 
The existence of an end-point is fairly securely
established by our physical argument. We have also made
sure that the the ratio of correlation lengths to the linear
system dimensions, 
$l_{ab}/L_{ab}$ and $l_c/L_c$, decrease rather than increase
as we pass the critical end-point with increasing $|\alpha_{2T}\eta|$. 
However, the general tendency of small system 
sizes to decrease discontinuities at the transition could have lead to 
an overall underestimate of the jumps due to finite size effects and
thus to an overestimate of $B_{lc}$.

The presence of sample disorder 
increases the value of $B_{lc}$, an effect which we report 
for random disorder in our simulation in Sec.\ \ref{sec:do3d}. 
Real YBCO samples exhibit in general not random disorder but  
clusters of oxygen vacancies and large scale sample 
inhomogeneities.
Oxygen clusters pin at low fields a considerable fraction of the 
vortex matter, so that the field can be divided as $B=B_{pinned}+B_{free}$. 
The fraction of pinned vortices will be increased by the presence of
sample inhomogeneities which provide 
regions at an effectively lower $\alpha_{T}$ that exhibit stronger pinning.
Such a decrease in the effective field of the 
free vortices ($B_{free}<B$) could considerably increase    
the observed $B_{lc}$.

Considering that the magnetic field at the end-point depends
so sensitively on the model parameters, critical fluctuations 
and disorder, the good 
quantitative agreement with the data of Ref.\ 
\cite{schilling_melt,schilling_anisotr} seems more than can be reasonably
expected from our model, and maybe is indeed a product of chance
in which $B_{lc}$ is in our simulation increased by inaccuracies 
of our model and/or finite size effects by the same amount 
that it is increased by disorder effects in the samples in  
Ref.\ \cite{schilling_melt,schilling_anisotr}.

\subsection{Beyond the critical end-point}
\label{sec:beycp}

It is often supposed that the  first order transition in YBCO
changes to second order below the end-point, where
no latent heat is visible but a ``step'' in the heat capacity $C$ remains
\cite{schilling_anisotr,Roulin_endpt}. We believe however that this 
``step'' can be identified with the onset of a small rounded peak in the 
superconducting specific heat $C_{s}=C-C_{n}$, ($n$ for normal state), 
which is known to arise from thermal fluctuations \cite{thouless}. 
In this section we shall examine experimental specific heat data
and numerical data in the light of this possibility.

\begin{figure}
\centerline{\epsfxsize= 6 cm\epsfbox{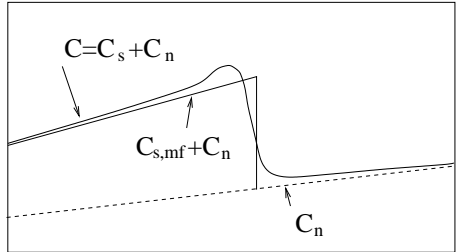}}
  \caption{Schematic view of a typical
        experimental specific heat curve without first order peak 
        (thick line) with normal state contribution (dashed line) 
        and mean-field value (thin line). }
  \label{fig:specht}
\end{figure}

The specific heat peak due to thermal fluctuations  
has been observed 
for example in niobium by Farrant and Gough \cite{Farrant&Gough},
where observations are 
in good agreement with theoretical predictions \cite{thouless}.
We find that the location
and the height of the peak as well as the  length of the rise 
(or width of the ``step'') in $\;C\;$ from  the low 
temperature value $C_{s, mf}$ 
($mf$ for mean-field) to its  maximum 
agree well for the niobium and YBCO measurements taken from Ref.\ 
\cite{Farrant&Gough} and \cite{schilling_anisotr}. 
We shall now  explain in detail how the data from both
references can be compared and give respectively values for 
the location, width and height of the specific heat peak or ``step''.
For a clearer picture of how the experimentally measured specific heat 
splits into normal and superconducting contribution
as well as how it compares to the mean-field contribution, a schematic
plot is given in Fig.\ \ref{fig:specht}.

Farrant and Gough give in Ref.\ \cite{Farrant&Gough}
the superconducting specific heat data already in terms 
of LLL scaling parameters. The plotted quantity is
$C_{s}/C_{s, mf}$.
The data in Fig.\ 3 from Ref.\ 
\cite{schilling_anisotr} which shows the specific 
heat in YBCO is given as  $C$ minus
$C(B\!=\!0)$. The latter near the ``step'' is approximately equal to the 
low temperature value $C_{s, mf}+C_{n}$, 
because for $B\! \rightarrow\! 0$, $\alpha_{T}\!\rightarrow\!-\infty$.
This approximate equality is also visible 
in Fig.\ 1 of the same  Ref.\ \cite{schilling_anisotr}. 
The plotted quantity,  $C-C(B\!=\!0)$,  is therefore approximately 
equal to $C_{s}-C_{s, mf}$. 

The peak in  $C_{s}$ in niobium obeys LLL scaling 
and is found at $\alpha_{T}\approx -7$.
The specific heat maximum in YBCO occurs at temperatures just above
the extrapolation of the first order transition line, 
which is located at $\alpha_{T}\approx -7.8$.
For the example curves for $B=0.25$T  and $B=0.5$ in 
Fig.\ 3 from Ref.\ \cite{schilling_anisotr}, the center of the 
broad specific heat peak is at $T\approx 91.4$K and $T\approx 90.7$K 
respectively, which both translate with the same previously used YBCO 
material constants to $\alpha_{T}= -7.2$. This is in 
very good agreement with niobium.

The width of the specific heat rise  in niobium $\Delta \alpha_{T}\approx 2$. 
For $B=0.25$T, no sharp step feature is visible in the YBCO data.
The specific heat rise from the low temperature value to the maximum 
takes place in the temperature region  
$91-91.4$K, which corresponds to $\Delta \alpha_{T}= 2.8$,
broader than in niobium. 
For $B=0.5$T one might suspect a step-like feature located
$90-90.5$K. This width corresponds to $\Delta \alpha_{T}= 2.3$,
a value in agreement with the niobium data. 

For niobium  $C_{s}$ is at its maximum 5\% larger than 
$C_{s, mf}$. In YBCO we have to divide  the plotted data by 
$C_{s, mf}$ to compare with this value. $C_{s, mf}$ is roughly given 
by the step in $C$ at the zero field transition, which we take from 
Figure 1 of Ref.\ \cite{schilling_anisotr} as  
$C_{s, mf} \approx$ 60 mJ/mole K$^2$. For all fields 
the specific heat ``step'' in YBCO is of the order of 
1.5  mJ/mole K$^2$, which gives 
$(C_{s}-C_{s, mf})/C_{s, mf}\approx 2.5$\%, 
which we consider as reasonable agreement with in niobium. 
Exact agreement cannot be expected for the following reasons. 
The value of $C_{s, mf}$ we used for YBCO is a rather 
crude approximation. For niobium $C_{s, mf}$ is also 
somewhat uncertain,
because it depends on the choice of $C_n$
(see Fig.\ \ref{fig:specht}), which is not directly
measurable but extrapolated from fits to higher 
temperature data.  Also the data in YBCO at fields below 
$B_{lc}$ are in a region where the LLL approximation is
inadequate.

\begin{figure}
\centerline{\epsfxsize=8cm\epsfbox{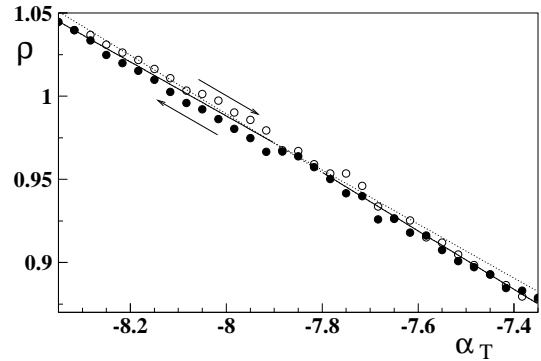}}
  \caption{Plots of  the order parameter density 
  $\rho$ upon heating and cooling for $|\alpha_{2T}\eta|=3$.
  Solid lines are linear fits for the regions above and below the 
   extrapolation
  of the first order transition line, extrapolated as dotted lines.
  $N_{ab}=72$, $N_{c}=80$.}
\label{fig:spechtsim}
\end{figure}

If the step feature is mainly due to the increase of LLL fluctuations, 
it should be observable in our simulation. A ``step'' in the
superconducting specific heat corresponds to a change in the slope
of the magnetization, in simulation terms a change of 
$\partial\rho/\partial\alpha_T$. Figure \ref{fig:spechtsim}
shows $\rho$ upon cooling and heating
for $|\alpha_{2T}\eta|=3$, well beyond the critical end-point.
The data for $\alpha_{T}<-7.9$, i.e.\ below the region where
the first order transition takes place in more layered samples, 
is very strongly affected by hysteresis. The sweep rate 
being of the same order as usual, this is a sign of very long 
fluctuation time scales. We shall come back to this point in Sec.
\ref{sec:opc}. Linear fits to the data for both cooling and heating 
below and above the extrapolated transition line,
$\alpha_{T}<-7.9$ and $\alpha_{T}>-7.8$, give a change in slope of 8\%,
larger than in the experimental data from both niobium and YBCO. 

In summary we have seen that 
the specific heat ``step'' in YBCO at different fields 
has approximately the same amplitude as well as width and position 
when expressed in terms of $\alpha_{T}$, 
i.e.\ the  ``step'' feature obeys LLL  scaling 
(LLL scaling has for the steps associated with the first order
transition previously been established in Ref.\ \cite{pierson}).
The corresponding data from our numerical simulation are 
not equilibrated and therefore not very accurate, but consistent 
with a similar ``step'' feature.
The semi-quantitative agreement between 
YBCO and niobium strongly suggests that 
we are dealing with the same phenomenon and therefore that there really 
is no sharp specific heat step in YBCO beyond the critical end-point.

\section{Order Parameter Correlations}
\label{sec:opc}
The existence of the critical end-point implies that no symmetries are  
broken at the transition, which means it cannot be a 
liquid-crystal transition. Investigation of the nature 
of order parameter correlations described in this section 
confirms that the state below the transition is a vortex liquid. 
However, neutron diffraction patterns corresponding to a triangular
lattice and an electrical resistance close to zero 
below the first order transition 
line suggest very long-range vortex correlations. 
We shall argue that the extremely 
fast growth of correlation 
length scales and relaxation time scales below the first order
transition, which have been  
theoretically predicted in the low temperature regime 
\cite{chin} as well as  observed in our simulation, can account 
for these effects.

\subsection{Static order parameter correlations}
\label{sec:sopc}
\begin{figure}
\centerline{\epsfxsize= 9 cm\epsfbox{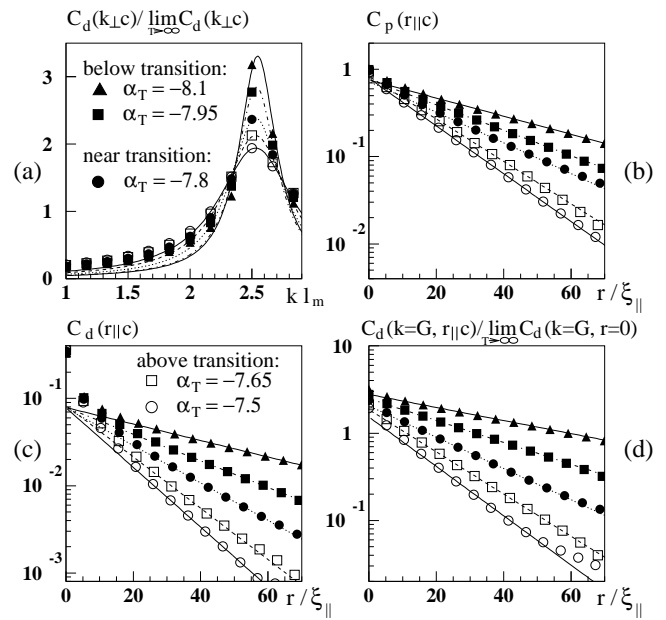}}
  \caption{Static order parameter correlations
             above and below the first order phase transition 
             at $|\alpha_{2T}\eta|$=1. 
            (a) Structure factor (with Lorentzian fits), 
            (b) phase correlations along the $c$-axis. 
             Plots (c) and (d) show density-density correlations along 
             the $c$-axis, where the correlator in (c) is in real 
             space for points with the same coordinates in
     	     the $ab$-plane, and the correlator in (d) is in
             Fourier space near the first reciprocal lattice 
             vector of the triangular lattice, $G$. 
             Note that in plots (b),(c),(d) the $y$-axes are
             such that parallel
             linear fits correspond to the same length scale. 
             The growth of length scales is slowest for the phase 
             correlations in (b) and fastest in (d).
             The system size is $N_{ab}$=72, $N_{c}$=270.
        }
  \label{fig:4}
\end{figure}

We have measured various different kinds of equilibrium 
order parameter correlation functions, including the structure 
factor as previously defined in Ref.\ \cite{dodgson}
and given  by the normalized density-density correlator in 
Eq.\ \ref{eq:cdnorm} (below) for $\Delta r=0$ along the $c$-axis. 
The peaks in the structure factor, which occur at the reciprocal
lattice vectors of the triangular lattice,  
reflect the crystalline correlations within each layer.
To examine order parameter correlations along the $c$-axis,
we have measured three different correlators. Firstly, we 
measured  the density-density correlations along a line parallel
to the magnetic field, $C_{d}(\Delta r\!\parallel\!c)$,
the same correlator that gave in Sec.\ \ref{sec:critpt} 
evidence of a diverging length scale at the critical end-point.
Secondly, we measured the decay of the $ab$-Fourier 
transformed density-density correlations, which are for
$\Delta r\!\parallel\!c=0$ the structure factor of the
system, depending on $\Delta r\!\parallel\!c$. Just as
for the structure factor \cite{dodgson} we normalize by the 
high temperature limit, so that the relevant quantity is
\begin{equation}
\label{eq:cdnorm}
C_d(k\!\parallel\!ab, \Delta r\!\parallel\!c)
/\lim_{\alpha_{2T}\rightarrow \infty}C_d(k\!\parallel\!ab, 0)\;,
\end{equation}
with 
\begin{eqnarray}
\lefteqn{C_d({\bf k}\!\parallel\!ab, {\bf \Delta r}\!\parallel\!c)=}
\nonumber&&\\ &&\;\;\;\;\;
\frac{1}{\langle|\psi|^{2}\rangle^{2}} 
   \left(\langle|\psi|^{2}({\bf k},\,{\bf r}) 
                       |\psi|^{2}({\bf -k},\,{\bf r}+{\bf \Delta
r})\rangle-\right.\nonumber\\
              &&\;\;\;\;\;\;\;\;\;\;\;\;\;\;\;\;\;\;\;\;\;\;\;
           \left.\langle|\psi|^{2}({\bf k},\,{\bf r})\rangle
               \langle|\psi|^{2}({\bf -k},\,{\bf r}+{\bf \Delta
r}\rangle\right)\nonumber .
\end{eqnarray} 
The definition of this correlator in its generalized, time dependent 
form in terms of the LLL coefficients and the high temperature limit 
are given in Sec.\ \ref{sec:3Ddc}, Eq.\ \ref{eq:cdofv3D} and
\ref{eq:defhtlim}. We find that this correlator
decays most slowly with  $\Delta r\!\parallel\!c$ for
$k_{ab}\approx G$, the first reciprocal lattice vector.
In addition to $C_{d}$ we also measured the phase correlator $C_p$ defined as  
\begin{equation}
\label{eq:cpdef}
C_{p}({\bf \Delta r}\!\parallel\!c)=\alpha_{T}\frac{\beta \langle
\psi^{\ast}({\bf r})\psi({\bf r\!+\! \Delta r})\rangle}{2 \pi \alpha_{H}}\;,
\end{equation}
which is expressed in terms of the LLL coefficients in Sec.\ \ref{sec:3Ddc}.
The prefactor is chosen such that that $C_p(0)$ is the
order parameter density $\rho$.
In Fig.\ \ref{fig:4} we show examples of static
density and phase  correlations above and below the transition
for $|\alpha_{2T}\eta|\!=1\!$, which corresponds to a transition 
temperature of 83 K in YBCO. 

Figure  \ref{fig:4}(a) shows examples of the 
structure factor with Lorentzian fits. 
The inverse width at half maximum, $\delta^{-1}$, 
of the peak near the first reciprocal lattice vector
is proportional to the crystalline correlation length $l_{ab}$
within one layer \cite{dodgson}, $\delta^{-1}=l_{ab}/2l_m$.
No qualitative change in
behavior is visible across the transition, just sharpening 
of the peak, which corresponds to 
an increase of $l_{ab}$. For the above  fits 
$l_{ab}/l_m=$2.66, 3.14, 3.88, 4.94, 5.46. This is shorter 
than the radius of the sphere $R/l_m=\sqrt{N/2}=6$ and 
shorter than the average 
distance between the 12 disclinations imposed by the spherical
geometry, $\sqrt{4 \pi R^2/12} \approx R$. 
This means that finite size effects on this data 
are not expected to be too large.

Figures \ref{fig:4}(b)-(d) show examples of measurements
of the three different correlators along the 
c-axis.  Figure \ref{fig:4}(b)
shows the  phase correlator, Fig.\ \ref{fig:4}(c) the real space
density-density correlator and  Fig.\ \ref{fig:4}(d) the
$ab$-Fourier transformed density-density correlator near
the first reciprocal lattice vector $G$.
In all cases we see an exponential decay of the correlation function
with a finite  length scale $l_{c}$ for density correlations
below as well as above the transition.
Only for a liquid phase would these correlation functions all have an 
exponential decay.  

All three correlators decay over half the system size $n<N_{c}/2$,
and then rise to $C(N_{c}d)=C(0)$ due to the periodic
boundary conditions. To extract decay length scales we fit to the 
first portion, approximately $\Delta r < (N_{c}d/6)$, of the decay,
as indicated by linear fits.
This region of decay is least affected by finite size effects, 
statistical noise and errors due to incomplete equilibration 
(see Sec.\ \ref{sec:numlim}).
The length scales extracted from the linear fits are of the same
order for all three correlators. Just above the transition the different
$l_c$ are almost equal. Below the transition the Fourier transformed
density-density correlations clearly have the largest length scale
with $l_{c,d}/\xi_{||}\!=\!58$ ($d$ for density) at $\alpha_T\!=\!-8.1$ and 
the phase correlator the smallest with $l_{c,p}/\xi_{||}\!=\!42$ 
($p$ for phase) at the same $\alpha_T$. 

At temperatures well above the transition we find the opposite behavior.
The density-density correlation length 
is at temperatures $\alpha_T \approx -4$ little more than half 
of the  phase correlation length, which 
agrees with the  simple high temperature expectation 
$\langle \psi(r)^{\ast} \psi(r')\rangle^2 \sim 
\langle |\psi(r)|^2 |\psi(r')|^2 \rangle$
and therefore for exponential decay $l_{c,p} \approx  2l_{c,d}$.  
However, because we 
find that below the transition $l_{c,d}$ is the longest 
and therefore dominant length 
scale, we shall in the following refer to the 
decay length scale of the correlator $C_{d}(k_{ab}=G)$ 
as $l_{c}$. 

\subsubsection{Temperature dependence of correlation lengths}
\label{sec:corrlen}

\begin{figure}
\centerline{\epsfxsize= 9 cm\epsfbox{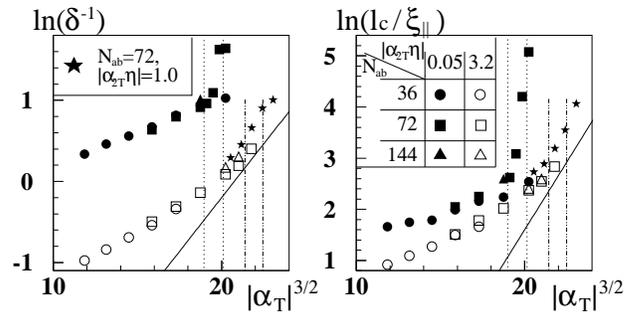}}
  \caption{Logarithm of correlation lengths plotted against  
           $|\alpha_T|^{3/2}$. Solid lines represent the 
           growth rate for the low temperature 
           regime as predicted  from perturbative  expansions 
           around zero temperature \protect\cite{chin}.
           Dotted and dot-dashed lines mark the width of the
           first order phase transition from the magnetization
           discontinuity for $|\alpha_{2T}\eta|$=0.05 
           and $|\alpha_{2T}\eta|$=1 respectively.
           Note the fast growth of length scales for 
           $|\alpha_{2T}\eta|$=1 below the phase transition. 
            Note also that
            the data below the transition for
            $|\alpha_{2T}\eta|$=0.05 is strongly affected by finite
            size effects ($l_{ab}\approx 10 l_m > R=6 l_m$). 
            } 
  \label{fig:statlen}
\end{figure}

In Fig.\ \ref{fig:statlen} we plot on the left the $\alpha_T$ 
dependence of the inverse width at half maximum of the first peak 
in the structure factor, $\delta^{-1}$, which is proportional to
the length scale of
in-plane crystalline order, $l_{ab}/l_{m}=2\delta^{-1}$,
and on the right the length scale $l_{c}$ as obtained from 
linear fits to the decay of density-density correlations.
The length scales have a discontinuity at the transition, 
which is found to grow with distance from the end-point,
as one would naturally expect. This discontinuity is clearly 
visible for $|\alpha_{2T}\eta|$=0.05. At large couplings, finite
size effects spread out the discontinuity.
The data for $|\alpha_{2T}\eta|$=1 shows a rapid growth of the  
length scales at and below the transition.

If the phase coherence of the Abrikosov state is examined 
in the presence of thermal fluctuations, one finds that in
and below three dimensions thermal fluctuations destroy 
phase coherence at any finite temperature \cite{moore-odlro}.
Under the assumption that there is only one relevant length scale 
describing phase order and crystalline 
order in the system,  perturbative studies for the 
continuum low temperature regime predict
an exponential growth of 
length scales with $|\alpha_{T}|^{3/2}$
as $l_{c}\!\propto\! \exp(A|\alpha_{T}|^{3/2})$ and 
$l_{ab}\!\propto\!\exp(0.5A|\alpha_{T}|^{3/2})$ \cite{chin}. 
The authors of Ref.\ \cite{chin}
estimate that $A$ may be given by its upper limit value $A=0.53$.
The slope of such growth behavior with $A=0.53$
is given by the solid lines in Fig.\ \ref{fig:statlen}.
The growth rate in our simulation data is at the lowest temperatures 
of the same order as the
theoretically predicted upper limit. However, the analytical result  
is from an expansion around zero temperature in the continuum limit. 
This regime is for numerical reasons described in Sec.\ 
\ref{sec:numlim} not accessible to our simulation. Therefore 
perfect agreement of our simulation results with this form cannot be
expected.

The data in Fig.\ \ref{fig:statlen} for  $|\alpha_{2T}\eta|$=3.2, 
which is beyond the 
critical end-point and has got no  phase transition, 
suggests that a faster growth of length scales 
sets in approximately where the phase transition is located in 
more layered samples. The reason why we have obtained no more data 
at lower  temperatures to confirm this tendency  
is just because of the fact that length 
and time scales (see Sec.\ \ref{sec:reltm}) grow so fast that we
found it impossible to equilibrate a sufficiently large 
system at even lower temperatures. (Remember that the same system size 
in units of $\xi_{||}$ corresponds to more layers for larger 
$|\alpha_{2T}\eta|$.) However, 
the  onset of fast growth of length scales can be expected from 
the fact  that  in the low temperature regime, 
where length scales are so long that the continuum approximation 
works well even for strongly layered materials, the data for different 
$|\alpha_{2T}\eta|$ must collapse onto a single curve and depend 
only on $\alpha_{T}$. At the transition, length scales in
the more continuous samples are in the appropriate scaling units 
distinctly shorter than for  low $|\alpha_{2T}\eta|$. 
For the length scales in these systems to approach the lengths
scales of the more layered systems, rather fast growth is necessary  
not far below the transition.

\subsubsection{The appearance of Bragg peaks}
\label{sec:bragg}
Just below the phase transition the correlation 
length scales in the simulation  
are for the range of coupling parameters 
that correspond to YBCO not comparable to the much larger 
length scales  
needed to give a signal in neutron diffraction experiments.
Although the structure factor in a 
liquid is rotationally symmetric, coupling with the
underlying lattice or preferred orientations given by twin planes 
may for long length scales lead to the 
appearance of Bragg-like peaks \cite{yeo-bragg}. 

Our simulation suggests that the vortex liquid is 
not far below the first order 
transition correlated and effectively crystalline
over length scales comparable to the system size or a
``Larkin''-like length scale  
(dependent on the amount of disorder present).
We can extrapolate the growth in length scales 
in Fig.\ \ref{fig:statlen}  
below the transition assuming the exponential growth rate estimated
in Ref.\ \cite{chin} and indicated by the solid lines.
For a decrease of $\Delta \alpha_{T} \approx 1.2$ for example,
which  corresponds in YBCO to cooling by only 1K or (1/4)K 
below the transition at 5T or 0.7T respectively, we obtain an
increase in $l_{ab}$  by a factor of 4 and an increase in 
$l_c$ by a factor of 16. At $\alpha_T \approx -10.5$ the crystalline 
correlation length $l_{ab}$ has according to the same estimate reached 
30 lattice spacings and $l_{c}$ the order of 10,000 $\xi_{||}$, which 
is for magnetic fields of $B\!\approx\;$5T or 0.7T  of the order of 
5,000 and 10,000 layers respectively. 
Correlation lengths of this order can be expected to 
lead in real samples to observable  neutron diffraction
peaks. 

\begin{figure}
\centerline{\epsfxsize= 8 cm\epsfbox{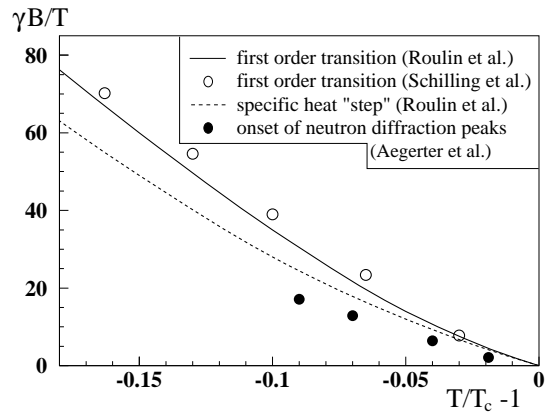}}
  \caption{Comparison of phase diagrams obtained on different 
          samples from  thermodynamic and neutron diffraction
          measurements.  All data is for $B \!\parallel\!c$.
          Empty circles and the solid line (extrapolated to $T_{c}$ 
          beyond $B_{lc}$) correspond to specific heat
          peaks. The dashed line represents the point of 
          steepest slope of the specific heat 
          ``step'' attributed to second order melting 
          in samples which do not show a  specific heat peak. 
          The data for the solid line, empty circles, the dashed line 
          and filled circles is taken from Ref.\ 
\protect\cite{Roulin_endpt,schilling_anisotr,Roulin_vgmelt,bham-neutscat} 
 respectively. 
          } 
  \label{fig:gammab}
\end{figure}

Figure \ref{fig:gammab} shows scaling plots of different 
experimental phase diagrams in different YBCO samples. 
We expect according to LLL scaling that the plots of 
$\gamma B(T)$ from samples
with a different mass anisotropy but otherwise identical
material parameters
collapse (see Sec.\ \ref{sec:simfot}).
Roulin {\it et al.}\ \cite{Roulin_endpt} report that the first 
order transition lines in different clean samples, 
represented by a solid line in Fig.\ \ref{fig:gammab},
as well as the  the line of steepest slope at the specific heat
step in disordered samples \cite{Roulin_vgmelt}, represented by a 
dashed line in Fig.\ \ref{fig:gammab}, collapse in this way. 
The solid and the  dashed line correspond to the specific heat peaks
in three different samples with different anisotropies each.   
The Schilling data (scaled using $\gamma=7.8$  
\cite{schilling_anisotr}) also agrees reasonably well with this 
scaling form. 

According to the vortex lattice melting picture, the points 
which mark the onset of neutron diffraction peaks in 
Fig.\ \ref{fig:gammab} (scaled using
$\gamma=4.3$ \cite{bham-neutscat}),
should coincide with the first order transition line. Comparison 
however shows that $\gamma B$ is roughly a factor 2 
below the first order transition lines. (In a scaling plot 
of $\gamma^2 B$, which should yield a collapse assuming
London-Lindemann-type melting, the discrepancy between the 
Aegerter and Schilling data would be even larger, roughly a factor 4.) 
Very substantial differences in other material parameters
would be necessary to make up for such a large deviation. 
In our picture the relative position of the two features is 
natural, because the onset of neutron diffraction peaks is attributed
to a crossover {\it below} the first order transition, 
when the neutron 
diffraction experiment becomes sensitive to the exponentially
fast diverging length scales.

However, one might say that comparing 
first order ``melting'' and onset of neutron diffraction 
is not a comparison of like with like, 
because the sample in the neutron diffraction experiment is
far too dirty to exhibit a first order phase transition. The authors
of Ref.\ \cite{bham-neutscat} estimate that their sample 
is comparable to  
samples from Ref.\ \cite{Roulin_vgmelt} and attribute 
the onset of neutron diffraction peaks to second
order freezing to a vortex glass. In this case they should
coincide with 
the line marking position of the steepest slope
of the specific heat steps in such samples,
which has been  interpreted as a second order 
vortex glass transition \cite{Roulin_vgmelt}. However, 
the plots in Fig.\ \ref{fig:gammab} show that this is
not the case.

Our scaling approach is based on the approximation 
$\alpha_{T} \sim 1/(\gamma B)^{2/3}$. Assuming that the 
first order transition line found by Roulin 
{\it et al.}\ \cite{Roulin_endpt} occurs as in our simulation at 
$\alpha_T \approx-7.5$, this means that the dashed line
and the onset of neutron diffraction in Fig.\ \ref{fig:gammab}, 
for which $\gamma B$ is reduced by a factor of about 3/4 and 1/2 
respectively, correspond to  $\alpha_T \approx -9$, and 
$\alpha_T \approx -10.5$ respectively.  
The $\alpha_T$ value for the onset of neutron diffraction peaks is 
in excellent agreement 
with our estimates 
as to  where $l_c\approx 10,000\xi_{||}$ and 
$l_{ab}\approx 30$ lattice spacings in the clean limit. 

\subsection{Relaxation times}
\label{sec:reltm}

The analysis of relaxation times is numerically difficult. 
It turns out that the dominant relaxation 
times are very large near the first order phase transition.
Accordingly thermal averaging is slow and measurements of the 
time decay 
of the density-density correlator have large statistical errors. 
However, although our data on relaxation times is of rather
poor quality, it is still of interest for comparison with the
analysis of the vortex dynamics in the 2D system as well as 
for comparison with non-equilibrium  measurements in YBCO.

As in our previous analysis of the 2D system \cite{kienappel}, 
we measure the relaxation time scales in the layered system
from the decay of the density-density correlator $C_d$
from Eq.\ \ref{eq:dencordef}
in its  Fourier transformed time-dependent form,
$C_d(k_{ab}, k_{c}, t)$.
We observe for high temperatures linear exponential decay of this 
correlator to zero for all $k$, where the $k_{ab}$ dependence of the decay 
time scales reflects, like in 2D\cite{kienappel}, the hexagonal 
order in the system. The time scales decrease monotonically 
with  $k_{c}$ for all
$k_{ab}$. For low temperatures, the time scales over which we measure
the decay are often much smaller than the longest decay time scales 
themselves, so that 
decay of the correlator is only observable over a fraction of its
initial value.  However, we can still, 
knowing that the vortex matter is  liquid, extract 
time scales by fitting the data assuming linear exponential decay. 
The longest time scales in the system are in all cases  given 
by the decay of the 3D Fourier component of $C_{d}$ at the first
reciprocal lattice vector in the $ab$-plane and $k$=0 along the 
$c$-axis.

\begin{figure}
\centerline{\epsfxsize= 9 cm\epsfbox{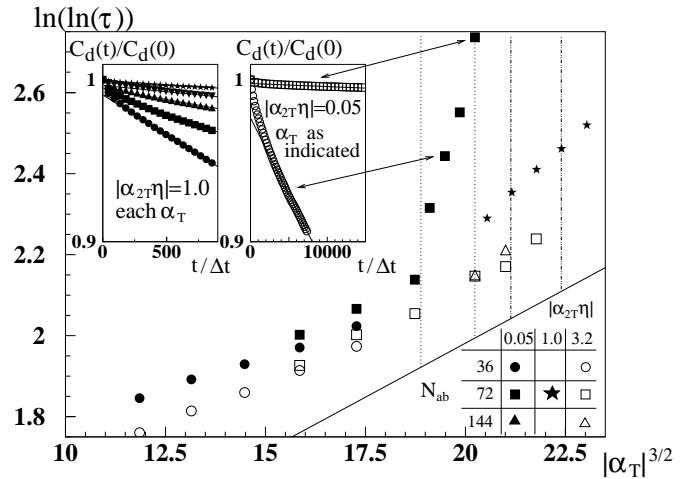}}
  \caption{Logarithm of logarithm of relaxation times 
           plotted against $|\alpha_T|^{3/2}$.
           Dotted and dot-dashed lines mark the width of the
           first order phase transition from the magnetization
           discontinuity for $|\alpha_{2T}\eta|$=0.05 
           and $|\alpha_{2T}\eta|$=1 respectively.
           The inset shows logarithmic plots of original 
           relaxation data with linear fits.
             } 
  \label{fig:3dtau}
\end{figure}

Figure \ref{fig:3dtau} shows a plot of the dominant 
time scales depending on $|\alpha_{T}|^{3/2}$ for different 
$|\alpha_{2T}\eta|$. The plots suggest very fast
growth behavior, roughly
\[\tau \sim \exp(c_1\exp(c_2|\alpha_{T}|^{3/2})),\]
where $c_1$ and $c_2$ are appropriate constants.
We expect an exponential growth of the dominant time
scales with the range of correlations, 
$\tau\!\propto\!\exp F l_{ab}^{\psi}$, if the dynamics is activated,
as it is in the 2D limit \cite{kienappel}. 
The data in Fig.\ \ref{fig:3dtau} together with 
the low temperature relation 
$\ln (l_{c})\propto 2\ln\l_{ab}\propto |\alpha_{T}|^{3/2}$ \cite{chin},
which is approximately also valid for our numerical data 
(see Fig.\ \ref{fig:statlen}), strongly suggests
such activated dynamics in the layered system. 

Figure $\ref{fig:3dtau}$ shows evidence that 
time scales increase discontinuously 
across the transition. If activated scaling of the type 
$\tau \sim \exp(c_1\exp(c_2|\alpha_{T}|^{3/2}))$ holds, 
the discontinuity in length scales should be amplified 
in the time scale discontinuity.
The increase in $\tau$ across the transition for 
$|\alpha_{2T}\eta|$=0.05
is roughly a factor 500. In the case $|\alpha_{2T}\eta|$=1
the time scales increase across the transition 
according to our fits  only by a factor 4.  
Although we expect the discontinuity in time scales to
decrease as $|\alpha_{2T}\eta|$ increases and the
end-point is approached, it is very likely that we 
strongly underestimate 
the time scales of exponential decay below the transition
for $|\alpha_{2T}\eta|$=1, because we fit to a very small regime of decay
(see inset of Fig.\ \ref{fig:3dtau}). As in the 2D case 
\cite{kienappel} the relaxation behavior at very early
times is faster than the final, linear exponential decay.

As in the case of length scales, the data  
for all $|\alpha_{2T}\eta|$ should collapse at low 
$\alpha_{T}$, i.e. in the continuum limit. 
This implies very fast growth of time scales 
below the transition $\alpha_{T}$ for continuous systems
with $|\alpha_{2T}\eta|$ beyond the end-point. This extremely
fast growth is reflected in the behavior of the system 
upon heating and cooling  for $|\alpha_{2T}\eta|$=3  
(see Fig.\ \ref{fig:spechtsim}), where strong hysteresis due to 
very slow decay of thermal fluctuations is observable 
below $\alpha_{T}=-8$.  

\subsubsection{Comparison with resistivity data}
Although the relevant time scales for dc resistance measurements in YBCO 
are non-equilibrium pinning time scales which are  not directly comparable
to the relaxation times measured in the simulation, some connections
can be made. While our time scales are irrelevant for flux
flow conductivity, they are relevant in a highly viscous vortex 
liquid regime, where the time scales
of thermal fluctuations or plastic deformations of the 
vortex structure are much larger then the pinning time.
In this case one speaks about a pinned vortex liquid \cite{blatterrev}. 
In such a liquid flux flow is strongly suppressed by pinning. 
When there is a large discontinuity of 
relaxation time scales at the first order transition, 
it is conceivable that the flux liquid becomes suddenly pinned 
and discontinuities in the resistive behavior occur.

A decrease of the discontinuity in time scales as 
one approaches the critical end-point  
agrees qualitatively well with the 
decreasing discontinuity in the resistive data 
as $B_{lc}$ is approached \cite{crabtree-rev}. 
The decrease and complete 
disappearance of the resistivity jump are difficult to 
explain with the assumption that the first order vortex lattice
melting changes to a vortex glass melting 
transition at $B_{lc}$. A change in resistive behavior and a 
discontinuity in the  resistivity at low currents
should remain for such a transition \cite{blatterrev}.

In experiment a very steep drop of resistivity to zero 
with decreasing temperature is 
observed even for $B<B_{lc}$, when there is no first order 
transition.  This feature can in our picture be attributed to 
the doubly exponentially fast divergence of decay time scales 
with decreasing temperature once  
the extrapolation of the transition line is crossed.
Although there is no time scale discontinuity, the system 
does a fast crossover from flux flow conductivity to 
a strongly pinned liquid regime.

\subsubsection{Irreversibility and magnetization measurements}
\label{sec:irrev}
A puzzling feature of the magnetization measurements near the 
first order transition in YBCO is the occurrence of
reversible-looking magnetization jumps without the corresponding
latent heat \cite{Junod_melt} which must according to the Clausius-Clapeyron 
relation (Eq.\ \ref{eq:clclp}) occur for an 
equilibrium measurement. In Ref.\ \cite{Junod_melt}
large magnetization jumps which would correspond to  
entropy jumps of up to 11 $k_{B}$/vortex/layer, 
more than 10 times higher than the largest jumps
observed in calorimetric measurements of approximately 
0.8 $k_{B}$/vortex/layer \cite{Roulin_endpt,schilling_anisotr},
are found without any evidence of a latent heat.

If magnetization and specific heat data do not obey the 
Clausius-Clapeyron relation, this must be due to lack of 
equilibration in the system. The connection between 
seemingly reversible features in the magnetization data 
and irreversibility has in some cases been shown 
experimentally.   For the vortex
transition in BSCCO Farrell {\it et al.}\ \cite{farrell} found that the 
magnitude of the magnetization jump is for standard SQUID measurement
techniques  strongly correlated to
irreversibility, even if irreversibility only becomes obvious at
temperatures {\it below} the magnetization jump.  
A similar effect in YBCO has been 
reported by Schilling {\it et al.}\ \cite{schilling-irrevmg},
where a change in slope of the seemingly reversible magnetization
is shown to be an effect of irreversibility. 

Very recent data in YBCO by Ishida {\it et al.}\ in the second
of Ref.\ \cite{ishida} shows magnetization jumps in a geometry 
$B\! \parallel\! a$ at $B=1.5$T, while in this geometry the 
specific heat jump has in similar samples been observed to 
disappear between $6$T and $4$T
\cite{schilling_anisotr}, a value in good agreement with 
our simulation data, if anisotropic scaling is assumed to apply.
In Ref.\ \cite{ishida} the ac ($f$=390 Hz) susceptibility is 
also measured and found to be an almost perfect image of the 
dc magnetization jump. This analogous behavior of the jumps in 
dc and ac susceptibility, of which the latter is clearly 
not a thermodynamic but a dynamic phenomenon due to non-equilibrium
pinning effects \cite{ishida}, is suggestive of irreversibility
effects even in the dc case. 

Although our simulation does not reproduce these effects, 
the relaxation time scales we measure give a clear indication 
why such irreversibility related effects can occur just
below the extrapolation of the first order transition line,
i.e.\ at $\alpha_{T}\!\leq\!-8$. 
Let us assume that the growth in relaxation time scales 
is of the type $\tau \sim \exp(c_1\exp(c_2|\alpha_{T}|^{3/2}))$ 
and further take an estimate of the slope of $\ln\ln\tau$ below 
$\alpha_{T}\!=\!-8$ from the low temperature data in 
Fig.\ \ref{fig:3dtau} for 
coupling $|\alpha_{2T}\eta|$=3.2 (beyond the end-point).
Using the estimate $\partial(\ln\ln\tau)/\partial 
(|\alpha_{T}|^{3/2})\approx 0.09$ 
let us consider a decrease in temperature from 
$\alpha_{T}\!=\!-8$ to $\alpha_{T}\!=\!-10$. 
This  corresponds to a temperature decrease of less than $0.3 K$ 
below the the extrapolation of the first order line 
at a field of $B\!\parallel \!c=0.25$T or, according to anisotropic
scaling with $\gamma\!=\!7$, to a field $B\!\parallel\! a$ of 1.75T.
The considered temperature interval is thus comparable to the width of the 
magnetization discontinuity in the data from Ref.\ \cite{ishida}.
For this decrease in $\alpha_T$ we can extrapolate an increase 
in time scales from  $\ln\ln\tau \approx 2.3$ to 
$\ln\ln\tau \approx 3.1$, which is by a factor $10^5$. For an increase
of time scales by a further  factor $10^5$ a further decrease of 
temperature by only $0.1 K$ is needed. 
This implies that cooling  by only fractions of a Kelvin below 
$\alpha_{T}=-8$  for fields $B<B_{lc}$ can cause the system to fall 
out of equilibrium. 

If one believes that falling out of equilibrium can lead to  
spurious magnetization jumps, then one may ask why a 
magnetization discontinuity is not always present, but in many cases 
disappears at low fields. For the data of Schilling 
{\it et al.}\ \cite{schilling_melt,schilling_anisotr} 
magnetization and specific heat data are found to be 
in good agreement with the Clausius-Clapeyron 
relation. A possible answer may be that the size of the  
magnetization discontinuity caused by  non-equilibrium effects depends
sensitively on the presence of sample disorder. Such effects may make 
it negligible  
against the thermodynamic magnetization discontinuity at the first 
order transition in some samples like  the ones used by Schilling 
{\it et al.} \cite{schilling_melt,schilling_anisotr}, but not in others.

\section{Effects of Disorder}
\label{sec:do3d}

We have investigated the effects of quenched
random disorder on the layered system. We have limited our study  
to point disorder, i.e.\ the disorder realizations 
were always uncorrelated in different layers. We shall describe  the 
effects of disorder on the first order transition as well as 
on correlations.

\subsection{Effects on the first order transition}
\vspace{-1cm}
\begin{figure}
\centerline{\epsfxsize=9cm\epsfbox{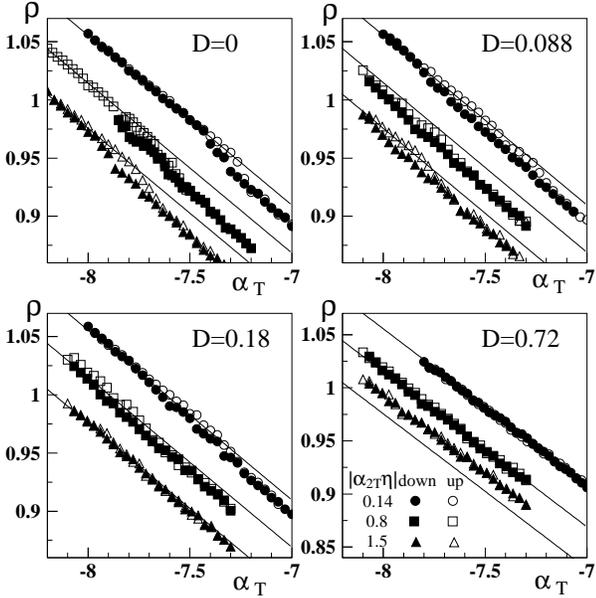}}
  \caption{Plots of  the order parameter density 
  $\rho$ upon heating and cooling for different values of disorder.
  Solid lines (identical for all $D$) are for easier comparison. 
  The hysteresis decreases with increasing disorder, slightly
  faster for high $|\alpha_{2T}\eta|$. 
   For $|\alpha_{2T}\eta|=0.14$ $N_{c}=20...16$ (increasing with $D$), 
   for $|\alpha_{2T}\eta|=0.8$ $N_{c}=60...40$ (decreasing with $D$), 
   and for $|\alpha_{2T}\eta|=1.5$ $N_{c}=60...50$  (decreasing with $D$). 
  For $|\alpha_{2T}\eta|=1.5$ $\rho$ is offset by -0.025.}
\label{fig:sweep_d}
\end{figure}

Figure \ref{fig:sweep_d} shows plots of  the order parameter
density $\rho$ upon heating and cooling for different values of 
disorder strength and coupling values varying from low coupling
($|\alpha_{2T}\eta|=0.14$) to values near the end-point 
($|\alpha_{2T}\eta|=1.5$).
We observe that the discontinuity persists in weak disorder, 
but decreases in amplitude and increases in width upon increasing
disorder until it disappears entirely. The statistical noise 
makes it difficult to determine the exact value of disorder
for which the discontinuity disappears. However, it looks  
as if for $D=0.18$ the  discontinuity near the end-point 
($|\alpha_{2T}\eta|=1.5$)
disappears, while for lower coupling
values it is still present. This behavior would suggest that in a
weakly disordered sample $B_{lc}$ can be higher 
than in the clean case; for the case $D=0.18$ we would expect 
$B_{lc}$ between $|\alpha_{2T}\eta|=1.5$ and 0.8, i.e.\ 
roughly between 2 and 5 Tesla in YBCO.  

Our simulation does not capture the existence of an
upper critical field $B_{uc}$. This can be interpreted as a 
sign that the appearance of $B_{uc}$ is strongly linked to the
presence of disorder clusters of a typical size, which are not  
present in a simulation with random Gaussian disorder. 
There is some experimental evidence for the link of $B_{uc}$ to 
the clustering of oxygen vacancies.  On the one hand it has been shown 
that the location of  $B_{uc}$ in YBCO is strongly correlated with the
field strength for which the fishtail effect occurs at lower 
temperatures \cite{newfeat}. On the other hand, the existence of the 
fishtail effect itself has been shown to be due to the 
presence of clusters of oxygen vacancies. A recent experiment 
by Erb {\it et al.}\ \cite{fishtail} demonstrates that for constant 
oxygen vacancy density the fishtail effect can be suppressed
by the destruction of vacancy clusters.

This picture can be supported by the following qualitative
argument. Let us assume the existence of a disorder length scale
$l_d$ and consider the relation of $l_d$ to $l_m$, the length scale 
which characterises the average vortex spacing and 
in the LLL also the order parameter coherence length. 
If  $l_m > l_d$, the system is 
only sensitive to a spatial disorder average, and the disorder 
can be considered as effectively weak. 
When $l_m < l_d$, the system is sensitive to the length 
scale of disorder, and disorder effects are not weakened. 
We can define a field  $B_{uc}\sim \Phi_{0}/l_d^2 $.
Then the disorder is effectively weak for $B < B_{uc}$, 
which is equivalent to the relation $l_m > l_d$.
In our simulation, disorder
is equally strong on all length scales, including length scales 
$l_d > l_{m}$ and may have effects similar to disorder clusters 
with a linear extent $l_d$. The disappearance of the first order
behavior in our simulation may thus be  qualitatively comparable to 
the disappearance of the first order
behavior observed in experiments at fields $B>B_{uc}$. 

\subsection{Effects on order parameter correlations}

By investigation of the order parameter correlations 
in the layered system near
and below values of $\alpha_{T}$ for which a first order transition
occurs in the clean system, we may gain some qualitative information 
about the low temperature vortex state where disorder has 
destroyed the transition, i.e.\ for $B_{lc, clean}< B < B_{lc}$, 
($B_{lc, clean}$ is the value of  $B_{lc}$ in the 
clean limit) or for $B>B_{uc}$. We choose to 
investigate the disordered system for a low coupling value, 
$|\alpha_{2T}\eta|=0.05$, for which we see very large discontinuities at 
the transition in the clean system (see Fig. \ref{fig:statlen} 
and \ref{fig:3dtau}).

\begin{figure}
\centerline{\epsfxsize=9cm\epsfbox{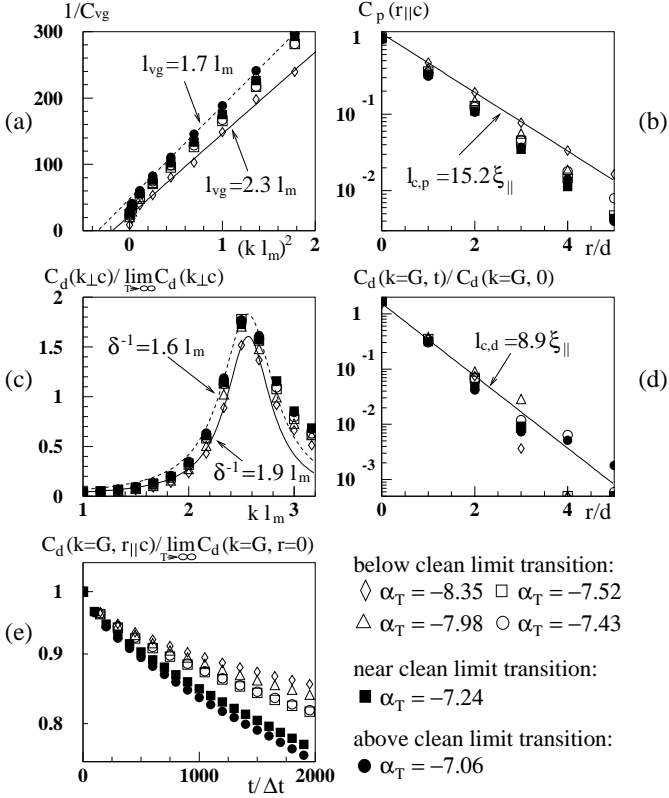}}
  \caption{Plots of  the order parameter correlations with disorder
   for  $|\alpha_{2T}\eta|=0.05$, $D=0.72$, $N_{ab}=72$, $N_{c}=30$:
   (a) vortex glass correlator, (b) phase correlations along the 
   $c$-axis, (c) structure factor, (d) density correlations along the 
   $c$-axis, (e) slowest time decay. All solid line fits are 
   for $\alpha_{T}=-8.35$, and the corresponding correlation lengths
   are indicated. In (a) and (c) the dashed lines are fits
   for $\alpha_{T}=-7.06$.}
\label{fig:do3d}
\end{figure}
To probe if the system is in a vortex glass state we
measure the vortex glass correlator \cite{2fisher&huse} defined as the 
disorder average over 
\begin{equation}
\label{eq:vgcorrdef}
C_{vg}({\bf r'})=\frac{\langle\psi^{\ast}({\bf r})
                       \psi({\bf r}+{\bf r'})\rangle\
                     \langle\psi({\bf r})
                       \psi^{\ast}({\bf r}+{\bf r'})\rangle}
                      {Q^4}\;,
\end{equation}
where $Q$ is defined in Sec.\ \ref{sec:nummod}.
$C_{vg}$ is a measure of the phase correlations in the system.
We can determine the Fourier transform of this correlator from
averages over the LLL coefficients as described in  
the Appendix, Sec.\ \ref{sec:2Ddc}. 
$C_{vg}$ is in a clean system trivially short-ranged. It is gauge 
dependent but has typically a Gaussian decay behavior,  
$\ln(C_{vg})\!\sim\! -r^{2}$ \cite{sasik&stroud&tesanovic}, 
which can for $D=0$ be observed in our simulation. 
In the presence of disorder, decay with a finite, temperature dependent
length scale is expected for a disordered vortex liquid 
\cite{2fisher&huse}.  
Exponential decay with a 
correlation length $l_{vg}$ in real space would appear as 
\begin{equation}
\label{eq:vgfit}
C_{vg}(k)\propto \frac{1}{k^{2}+l_{vg}^{-2}}.
\end{equation}

In Fig.\ \ref{fig:do3d} we show measurements of different
density and phase correlators at a disorder strength 
for which all signs of a first order phase transition have disappeared
from the heating and cooling measurements (not shown). In
Fig.\ \ref{fig:do3d}(a) $1/C_{vg}$ is plotted against $k^{2}$.
The behavior is in the plotted regime linear for $(k l_m)^2>0.25$.
The data for wave vectors $(k l_m)^2<0.25$ corresponds to correlations 
more than a third of the circumference of the sphere. 
It is strongly  affected by finite 
size effects and is therefore not taken into account
for linear fits according to Eq.\  \ref{eq:vgfit}.
The linear fits in Fig.\ \ref{fig:do3d}(a)   
have rather large errors. Firstly insufficient averaging over 
disorder configurations causes statistical noise.
Secondly and more importantly the intercept $-l_{vg}^{-2}$
depends sensitively on the exact choice of the linear fitting regime.
It is however visible that for any choice it is finite and
decreases with increasing $|\alpha_{T}|$. 
Although $l_{vg}$ grows with decreasing
temperature, it is in the measured temperature regime 
only of the order of a lattice spacing.
No long-range phase correlations typical for a 
vortex glass\cite{2fisher&huse} are visible.

The correlators we plot in Fig.\ \ref{fig:do3d}(b), (c), (d) and 
(e) have been introduced in Sec.\ \ref{sec:opc}. 
In Fig.\ \ref{fig:do3d} filled symbols correspond to temperatures just above  
and at the location of the first order transition in the clean 
system and empty 
symbols to temperatures below it. In none of the static correlators 
is there any rapid change of behavior across the location 
of the first order transition in the clean limit. 
However, in the relaxation behavior in Fig.\ \ref{fig:do3d}(e) 
an accelerated slowing down reminiscent of the discontinuity in the
clean limit is still visible.

All correlation lengths increase with $|\alpha_{T}|$. 
The correlation lengths $l_{ab}=2\delta^{-1}l_{m}$ and 
$l_{c}$ from density as well as from phase correlations
are reduced, for low temperatures by up to an order of magnitude, 
compared to the clean system (see Fig.\ \ref{fig:statlen}).
Unlike in the clean system,  $l_{c,p}$, i.e.\ the decay length
of phase correlations along the $c$-axis, is distinctly larger
than $l_{c,d}$, the density-density correlation length. The
phase correlation length in the plane from a fit
of the vortex glass correlator at $\alpha_{T}\!=\!-8.35$
is of the same order as the density-density correlation length 
at the same temperature. Although  $\alpha_{T}\!=\!-8.35$ is
distinctly below the phase boundary in the clean system, 
no signs of long-range phase correlations expected for a vortex glass
phase are visible. 

We have not attempted a quantitative analysis of relaxation 
behavior in the layered system plotted in Fig.\ \ref{fig:do3d}(e), 
because we have 
not monitored the decay behavior over long 
enough time time intervals to see the dominant decay rates. 
However, we expect a behavior which is qualitatively similar to the   
the 2D limit of the disordered vortex liquid. 
We give in the following an analysis of this relaxation behavior, 
which extends and partly supersedes our earlier analysis 
given in Ref.\ \cite{kienappel}. We include with Fig. 
\ref{figure2} a quantitative 
correction to data from Ref.\ \cite{kienappel} which was affected 
by an error in the simulation code.
\vspace{-0.3cm}
\begin{figure}
\centerline{\epsfxsize=  7.5cm\epsfbox{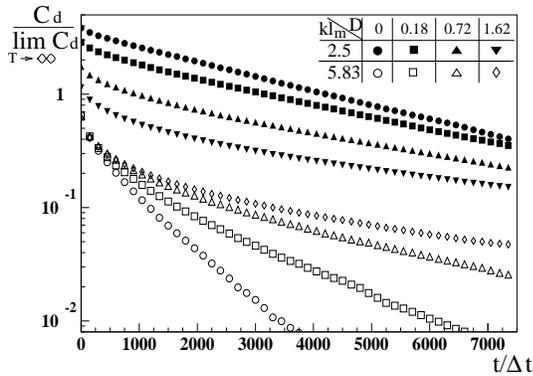}}
  \caption{Typical relaxation behavior
           with increasing disorder at $kl_m=2.5$ (near the first 
            reciprocal lattice vector) and for an arbitrary different 
           value of $k$. $N_{ab}=72$ and $\alpha_{2T}=-9$.}
  \label{figure3}
\end{figure}

Examples of the relaxation behavior of $C_d(k, t)$ with 
and without disorder are shown in Fig.\ \ref{figure3} 
for the 2D limit. 
The relaxation behavior in samples with disorder
is richer than our earlier analysis \cite{kienappel} suggests. A
seemingly faster decay than in the clean sample is visible at short times. 
However, this faster decay characterises only 
the initial decay behavior, which becomes clear in the
data in  Fig.\ \ref{figure3} for the case of strong disorder. 
At late times slow exponential
decay on a  time scale $\tau_{f}$ due to pinning to the disorder
becomes visible. To distinguish the
two regimes of decay, we define an initial relaxation time 
$\tau_{in}$ by $\tau_{ in}=\tau_{1/2}/\ln(2)$, where 
$\tau_{1/2}$ marks the time when the correlator has
decayed to half its initial value.
The final time scale $\tau_{f}$ can be found by 
exponential fits to the decay 
at late times. 
\vspace{-0.3cm}
\begin{figure}
\centerline{\epsfxsize= 7.5cm\epsfbox{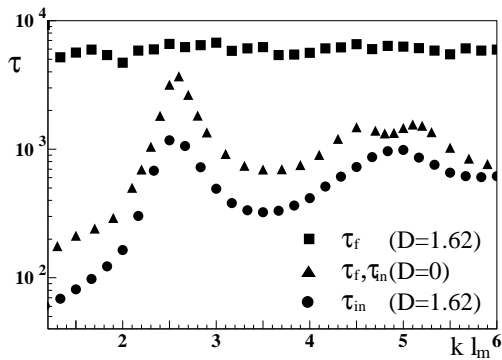}}
  \caption{Initial and final relaxation time scales of the 
  decay of the density-density correlator in reciprocal space  
  with and without disorder.  For $D\!=\!0$ system size $N_{ab}\!=\!200$, 
  for $D\!=\!1.62$ $N_{ab}\!=\!72$. $\alpha_{2T}=-9$.}
  \label{fig:tauofk-dn}
\end{figure}
Figure \ref{fig:tauofk-dn} shows a typical logarithmic plot of 
the $k$ dependence of initial and final relaxation times 
in the case of disorder compared to the clean case at the same 
temperature. The initial relaxation
times $\tau_{in}$ in the disordered case show qualitatively the same
$k$ dependence reflecting the
hexagonal structure as in clean samples\cite{kienappel}, while the final
relaxation times $\tau_{f}$ do not depend on $k$. This seems
natural, because the disorder contributions are
on average equally strong for all  $k$. 
\vspace{-0.3cm}
\begin{figure}
\centerline{\epsfxsize=8.5cm\epsfbox{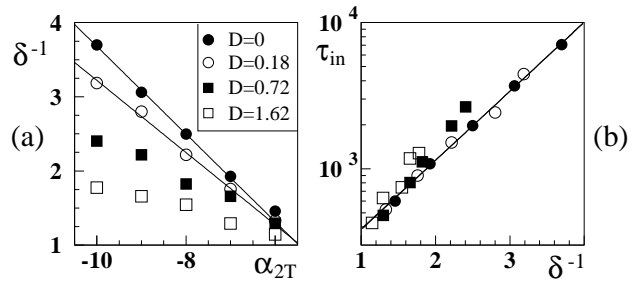}}
  \caption{ (a) The inverse width of the first peak in the structure factor 
   with linear fits for no and weak disorder in the low temperature
  regime.  For $\alpha_{2T}=-10$ and  $D=0$ $N_{ab}=200$, otherwise 
    $N_{ab}=72$.   
    (b) Data of length and time scales plotted as $\ln(\tau_{in})$ versus
    $\delta^{-1}$. With linear fits for $D=0$ and $D=0.18$.}
  \label{figure2}
\end{figure}
The dependence of the crystalline correlation length 
on $\alpha_{2T}$ is plotted for different disorder strengths 
in Fig.\ \ref{figure2}(a) and the dependence of the 
initial relaxation times on this length in Fig.\ \ref{figure2}(b). 
The initial relaxation times in the disordered system follow 
like in the clean system activated dynamics \cite{kienappel}, in 
which the dominant relaxation time scale grows linearly with 
the crystalline correlation length.

The relaxation behavior in the disordered layered system 
plotted in Fig.\ \ref{fig:do3d} looks similar to the 
2D behavior at early times; the initial relaxation is faster 
than in the clean case. 
A more quantitative analysis of the relaxation behavior 
in layered systems would be
of some interest in comparison to resistivity data in disordered 
samples. An analysis of the resistive behavior in YBCO before and
after artificial introduction of point defects by electron irradiation 
\cite{fendrich} concluded that transport in the disordered vortex
liquid is dominated by viscosity rather than single vortex pinning,
i.e.\ that the equilibrium relaxation time scales in simulations
without a driving field could be relevant time scales. In the
resistive measurements after irradiation in Ref.\ \cite{fendrich} 
no signs of a continuous vortex glass transition replacing 
the first order behavior before irradiation have been found.
This absence of any thermodynamic phase transition in the case
of strong disorder  is in agreement with our simulation results.

\section{Discussion and Summary}
\label{sec:sum-dis}

In our numerical work we have found a first order transition 
which on the one hand coincides with the first order transition 
in YBCO, but on the other hand is not associated with vortex lattice
melting. 
An analysis of what triggers the decoupling transition at the
particular $\alpha_{T}$ where it is observed is beyond the scope of
this paper. The answer to this question could lie in 
the observation made by Pierson and Valls \cite{pierson} that the 
first order transition in YBCO coincides with the onset 
of 3D LLL fluctuations. 
Because even strongly layered systems are 3D in nature 
below the first order transition, the onset of 3D LLL fluctuations
can be expected to occur at a fairly constant value of 
$\alpha_{T}$, and this may be reflected in approximately 
constant $\alpha_{T}$ along the phase transition line. 

Some readers may believe that the nature of the
first order transition  in YBCO is vortex lattice melting and 
dismiss the absence of a vortex lattice state 
below the transition in our simulation
as an artifact of the spherical boundary conditions 
we use. In this case however it would seem
astonishing that this first order transition associated with 
vortex lattice melting should quite happily persist in our 
simulation in the absence of a vortex lattice. Besides, 
a first order vortex decoupling transition \cite{wilkin&jensen} 
and another similar \cite{footnote}
thermodynamic transition \cite{nguyen_new}, both 
not associated with vortex lattice melting, have also been observed in 
simulations of different models using periodic boundary conditions. 

\subsection{Comparison with previous LLL-LD simulations}
From previous simulations using the same model, but with
periodic boundary conditions, a single vortex lattice 
melting transition is reported \cite{sas&str3D,hu&mcd3D}.
We find there is a disagreement in the location of the transition 
between our simulation and Ref.\ \cite{hu&mcd3D} for weak couplings.
This is certainly due to  the different choice of boundary
conditions, because in the the 2D case the two choices of
boundary conditions are known to yield different results 
\cite{dodgson,o'neill&lee,hu_macdonald&kato_nagosa}.
As the coupling increases, 
the disagreement in the location of the transition  
between our results and those from Ref.\ \cite{hu&mcd3D}
disappears. For couplings high enough to see the critical end-point, 
the LLL-LD model has never been investigated using 
periodic boundary conditions. 

The LLL-LD model simulations using periodic boundary conditions exhibit a 
vortex lattice state below the transition.
If the transition was of the same nature for periodic boundary 
conditions as for spherical layers, the apparent 
vortex lattice state could be accounted for 
by the use of system sizes smaller than the correlation 
lengths. (The largest system 
sizes in these simulations were of the order 
of only 40 vortices $\times$ 20 layers.) 
Small system sizes together with the restrictions in  degrees of freedom 
imposed by periodic boundary conditions 
may make a vortex liquid with very long length scales 
indistinguishable from a vortex lattice.

\subsection{Relevance for BSCCO}

We have so far only compared our numerical data  with experiments 
in YBCO. Near the phase transition in BSCCO neither 
the LLL approximation nor the LD model are valid. 
The phase diagram of BSCCO has a first order transition line \cite{zeldov}, 
which occurs at much smaller applied
magnetic fields than in YBCO. Near the BSCCO transition
$H/H_{c2}$ is of the order $10^{-3}$. This
means that the LLL approximation we use in 
our simulation cannot be expected to apply.
The strongly anisotropic character of BSCCO
is such that Josephson coupling between the layers may be 
negligible compared to  electromagnetic coupling effects 
\cite{blatter-emcoup} and so BSCCO is not accurately described
by the LD model.

However, some qualitative points of comparison
can be made. It is for example noteworthy that in BSCCO 
the material parameters $\kappa$ as well as the 
layer periodicity $d$, $T_{c}$ and 
$\partial B_{c2}/\partial T$ are of the same order as in YBCO, but 
typical estimates of the mass anisotropy $\gamma$ 
are between one and two orders of magnitude larger.
This means that the end-point of our numerical transition line
translates via the relation $B_{lc} \propto 1/\gamma^4$ to 
fields that are between 4 and 8 orders of magnitude lower than in YBCO.
An experimental observation of a lower critical end-point is therefore 
not to be expected in BSCCO. 

In the low temperature limit, where all length scales are large, 
both YBCO and BSCCO should show the same universal behavior.
Should our phase diagram be valid so that there 
is no thermodynamic vortex lattice melting transition at a finite 
temperature below the experimentally observed first order transition 
in YBCO, then BSCCO
should also be in a vortex liquid state below the first order 
transition. The consequence that the first order transition in BSCCO
is not of a genuine vortex lattice melting character is in agreement
with recent experimental evidence that hexagonal neutron diffraction
patterns, which signify that a vortex lattice or a vortex liquid with very
long length scales, can be observed above as well as below the 
first order transition line \cite{forgan}.

\subsection{Summary}
We have numerically calculated the phase diagram of a layered
superconductor and found a first order transition line between two
vortex liquid states. The length scales of order
parameter correlations parallel and perpendicular to the magnetic
field as well as the longest relaxation  time scales increase 
discontinuously at the transition but remain finite as the temperature 
is lowered. 
As the coupling strength between layers increases,   
the discontinuities in length and time scales decrease until 
the transition line ends at a critical end-point. Shape, location
and anisotropic scaling properties 
of the  transition line and its end-point as well as the size of the 
reversible magnetization discontinuity are in excellent agreement with 
the experimental first order transition line and its low field end-point
in YBCO. The approximate location of the end-point can be predicted from a 
qualitative argument assuming that the transition is of a 
layer decoupling nature. However, the exact
quantitative agreement of the location of the end-point with
experiments by Schilling {\it et al.}\ \cite{schilling_melt} could
be a chance result due to cancellation of inaccuracies of 
our numerical model at low fields and/or finite size effects
with disorder effects in real samples. 

Our results 
suggest that the transition in YBCO, which is commonly interpreted 
as vortex lattice melting, is of a liquid-liquid nature, 
with a low temperature vortex liquid phase in which  
length scales grow exponentially
fast and time scales due to activated dynamics doubly  
exponentially fast with decreasing temperature. 
We argue that because not far below the first order 
transition the vortex liquid is highly viscous and  effectively crystalline
over large length scales, our picture can account for many 
experimental features including resistance drops and Bragg peaks,
which have so far been taken as evidence for the vortex lattice 
melting scenario.

We have investigated the effect of quenched random point disorder 
on the system and found that the first order transition can be suppressed
by strong disorder and that the low field critical end-point can be 
shifted to higher fields by weak disorder. We find no evidence of a 
thermodynamic transition to a Bragg glass phase or 
vortex glass phase replacing the first order transition 
line in the presence of strong disorder.
 
\section*{acknowledgements}
We would like to thank Sai-Kong Chin 
and Matthew Dodgson for useful interactions. 
AKK acknowledges financial support from a 
Manchester University 
Research Studentship and EPSRC.

\appendix

\section{Details of the Hamiltonian}
\subsection{Quartic energy term}
\label{sec:quarten}
The quartic coupling term in Eq.\ \ref{eq:simham2} can 
in each layer be expressed in
terms of the LLL eigenfunctions in the following way
(we omit the layer numbering indices):
\begin{equation}
\label{eq:bfsq}
\frac{{\cal H}_{quartic}}{k_{B}T}=\frac{1}{k_{B}T}\; \;
\frac{\beta_{2D}}{2}\;d_{0}\int d^{2}r\, |\psi|^{4}
=\frac{1}{2N}\sum_{p=0}^{2N}|U_{p}|^{2},
\end{equation}
with $U_{p}=\sum_{m=\max(0,p-N)}^{\min(p,N)}\,f(m,p-m)\,v_{m}v_{p-m}$,
with
$f(m,n)=A_{m}A_{n}\,(B(m+n+1,\,2N-m-n+1))^{1/2}$, where
$A_{n}$ as defined in Eq.\ \ref{eq:lll} and $B$ is the beta function.

\subsection{Coupling to disorder}
\label{sec:append2}
Let the Gaussian random disorder $\Theta({\bf r})$ in each  layer 
be expanded as \[\Theta=\frac{k_B T}{d_0 Q^2}
\sum_{l=0}^{\infty}\sum_{m=-l}^{l} a_l^m \tilde{Y}_{l}^{m},\]
where the $\tilde{Y}_{l}^{m}$ are spherical harmonics normalized with 
respect to the spherical surface of radius $\sqrt{N/2}$.
For real $\Theta$, $a_{l}^{m \ast}=a_{l}^{-m}$.
The $a_l^m$ are chosen independently from a 
random Gaussian distribution
with zero mean and variance $\overline{|a_l^m|^2}=D$.
Calculating the disorder correlations 
$\overline{\Theta({\bf r})\Theta ({\bf r}')}$ and 
inserting $\Delta$ as defined in Eq.\ \ref{eq:deltadef} 
fixes $D=(d_0 Q^{2}/k_{B}T)^{2}\Delta$.
For the integration of (layer indices omitted) 

\begin{eqnarray}
\label{eq:discont}
\frac{{\cal H}_{dis}}{k_B T}
 &=&\frac{d_{0}}{k_B T}\int d^{2}r \,\Theta({\bf r})|\psi({\bf r})|^2 
 \\
  &=& \sum_{p,q=0}^N \sum_{l=0}^{\infty} \sum_{m=-l}^{l}
     \int d^{2}r\;\; a_l^m 
\tilde{Y}_{l}^{m} v_p^\ast\phi_p^\ast v_q\phi_q \; . \nonumber
\end{eqnarray}
we define for $0\leq m\leq l$ and $0\leq q\leq N-m$
\begin{eqnarray}
\label{eq:ilmndef}
\lefteqn{I_{l,q}^{m}
=\int d^{2}r \tilde{Y}^{m}\phi_{q+m}^{\ast}\phi_{q}}&&\\
&=&\!A_{q}\!A_{q\!+\!m}\sqrt{\frac{(2l\!+\!1)(l\!+\!m)!}{2\pi N(l\!-\!m)!}}
             \;\;\frac{(-1)^{m}}{m!}B(N\!-\!q\!+\!1,q\!+\!m\!+\!1)\nonumber\\
&&\times\, _{3}F_{2}(
             m\!-\!l,\, m\!+\!l+\!1,\,q\!+\!m+\!1;\;\;
             m\!+\!1,\,N\!+\!m\!+\!2;\; 1)\nonumber\;,
\end{eqnarray}
where $_{3}F_{2}$ is a generalized hypergeometric function. 
Substituting these integrals into Eq.\ \ref{eq:discont}, 
truncating the contributions of noise components 
with  frequencies $l>l_{\max}=N$ 
(for a standard system size of $N=72$ these are more than 
$10^{20}$ times smaller than the largest contributing terms),
and defining
\[g_{q,p}=\sum_{l=p-q}^N a_l^{p-q} I_{l,q}^{p-q}\]
yields 
\begin{equation}
\label{eq:hdis}
\frac{{\cal H}_{dis}}{k_B T}
=\sum_{q=0}^N \sum_{p=q}^N 
         g_{q,p} v_{p}^{\ast} v_{q} + \delta_{q,p} \times c.c. \,.
\end{equation}
Note that once  $g_{q,p}$ is calculated 
the original coefficients $a_l^m$ can be discarded
and need therefore not all simultaneously be held in memory.

\section{Correlations}

To compute correlations in reciprocal space, 
we perform the spherical equivalent of a Fourier transform, 
the expansion in the discrete set of normalized spherical 
harmonics, $\tilde{Y}_{l}^{m}({\bf r})$. 
To a value of $l$ corresponds $k\!=\!l/R$. Because 
the liquid is isotropic, the correlator in $k$ space 
depends only on the magnitude of $k$, i.e.\ only on $l$ and not on $m$.
For better averaging we calculate the correlator for all $m$
and average over the different $m$.  

\subsection{Vortex glass correlations}
\label{sec:2Ddc}
We calculate the Fourier transform of the correlator
as defined in Eq.\ \ref{eq:vgcorrdef} in terms of thermal
averages of the LLL coefficients. 
\begin{eqnarray*}
\lefteqn{C_{vg}(l/R)=\frac{1}{Q^4(2l\!+\!1)\,4 \pi R^2}\int\! d^{2}r\,
              \int\! d^{2}r' }&&\\
     &&\sum_{m=-l}^{l}\tilde{Y}_{l}^{m}({\bf r}) 
               \tilde{Y}_{l}^{-m}({\bf r'})
     \left\langle\psi^{\ast}({\bf r})\psi({\bf r'})\right\rangle
     \left\langle\psi({\bf r}) \psi^{\ast}({\bf r'})\right\rangle\\
    &=&\!  \sum_{m=-l}^{l} \;\;
        \sum_{n,n'=\max(0,-m)}^{\min(N,N-m)}
        \frac{ I_{l,n}^{|m|} I_{l,n'}^{|m|} 
        \langle v_{n+m}^{\ast} v_{n'+m} \rangle 
        \langle v_n v_{n'}^{\ast} \rangle}{2\pi N(2l\!+\!1)}\;,
\end{eqnarray*}
where the $I_{l,n}^{m}$ are defined in Eq.\ \ref{eq:ilmndef}.

\subsection{Density-density correlations along the $c$-axis}
\label{sec:3Ddc}
The real space density-density correlator for $\Delta t=0$
and $\Delta {\bf r}\!\parallel\!\hat{c}$, $\Delta r\!=\!nd$
is given by 
\[C_{d}({nd})=\frac{\langle|\psi({\bf r})|^{2}|\psi({\bf r}+
 nd \hat{c})|^{2}\rangle}{\langle|\psi|^{2}\rangle^{2}}-1.\]
Taking the spatial average over all $\bf{r}$ in one layer
involves the same integrals over LLL eigenfunctions as 
the calculation of the quartic energy term. Together with 
spatial averaging over different layers this yields

\begin{equation}
\label{eq:cdofnddef}
C_{d}({nd})=\frac{\sum_{p=1}^{N_{c}}\left \langle \sum_{q=0}^{2N_{ab}}
             \tilde{U}^{n \, \ast}_{p,q}
             \tilde{U}^n_{p,q}\right\rangle}{ 
            \sum_{p=1}^{N_{c}} \left \langle  \sum_{q=0}^{N_{ab}}  
         v_{p,q}^{\ast}v_{p,q}\right\rangle^2} -1\,,
\end{equation}
where 
\[
\tilde{U}^n_{p,q}=
\sum_{m=\max(0,q-N_{ab})}^{\min(q,N_{ab})}\,f(m,q-m)\,v_{p,m}v_{p+n,q-m},
\]
with $f(m,q-m)$ as defined in Sec.\ \ref{sec:quarten}.

The 2D density-density correlator in k space from 
reference \cite{dodgson}
is easily generalized to three dimensions as  
\begin{eqnarray}
\label{eq:cdofv3D}
\lefteqn{C_d(l/R,nd,\Delta t)= \frac{2\pi N_{ab}}{ (2l\!+\!1)} 
         \sum_{p=1}^{N_{c}}\left \langle  \sum_{q=0}^{N_{ab}}  
         v_{p,q}^{\ast}v_{p,q}\right\rangle^{-2} \times }\!&& \nonumber\\
       &&
          \sum_{m=-l}^{l}\sum_{p=1}^{N_{c}}  \left\langle 
              \sum_{q=\max(0,-m)}^{\min(N_{ab},N_{ab}-m)} 
        \!\!\!\!\!   v_{p, q+m}^{\ast}(t)v_{p, q}(t)I_{l,q}^{|m|}\right.
            \times \nonumber \\ 
         && \left.
       \sum_{q'=\max(0,-m)}^{\min(N_{ab},N_{ab}-m)}
        \!\!\!\!\!    v_{p+n,q'+m}(t')v_{p+n,q'}^{\ast}(t')I_{l,q'}^{|m|}
              \right\rangle _{c} \, ,
\end{eqnarray} 
where $c$ signifies the connected average and the $I_{l,n}^{m}$ are 
defined in Eq.\ \ref{eq:ilmndef}.
The high temperature limit of this correlator is for $t\!=\!n\!=\!0$
easily calculated analytically \cite{dodgson} as 
\begin{equation} 
\label{eq:defhtlim}
\lim_{\alpha_{2T}\rightarrow
\infty}C_d(l/R,0,0)=\frac{(N!)^2}{(N-l)!\;(N+l+1)!}\;.
\end{equation} 

For analysis of relaxation times this correlator is 
numerically Fourier transformed in the $c$-direction. 
To make $C_d(l/R,nd,t)$ converge 
in the continuum limit we choose $\xi_{||}$ as unit of length when 
integrating along the $c$-axis:
\begin{equation} 
\label{eq:fpcorft}
C_{d}(l/R,q,t)=\frac{1}{\sqrt{\eta}}\sum_{n=1}^{N_{c}}
C_d(l/R,nd,t) \cos(q\!\times\! nd)
\end{equation} 
where q takes values $q=2\pi m/(N_{c}d)$  for $m=0...N_{c}/2$.

\subsection{Phase correlations along the $c$-axis}
The phase correlations along the $c$-axis as defined in Eq.\
\ref{eq:cpdef} are easily expressed
in terms of the LLL coefficients using the orthonormality of
the LLL functions and the identity 
$\alpha_{T}=2 \pi \alpha_{H}/\beta Q^2$:   
\begin{eqnarray}
\label{eq:cc}
C_{p}(nd)&=&\frac{\alpha_{T}\beta}{2 \pi \alpha_{H}}
  \left \langle \frac{1}{4\pi R^2 }\int d^2r \;
      \psi^{\ast}(r)\psi(r+nd) \right\rangle\\
      &=&\frac{\alpha_{T}}{\alpha_{2T}}\times \frac{1}{2\pi N_{ab}N_{c}}
        \left \langle 
          \sum_{p=0}^{N_{ab}}\sum_{q=1}^{N_{c}} v_{p,q}^{\ast} v_{p,q+n}
        \right\rangle \nonumber.
\end{eqnarray}

\section{Finite size effects}
\label{sec:numlim}

The limitations in the region of parameter space for which we can 
run our simulation as well as the limitations in accuracy of our 
measurements are mainly due to limited 
availability of processor time.
The simulation time grows, as seen in Sec.\
\ref{sec:reltm}, roughly doubly exponentially fast with decreasing
$\alpha_{T}$. 
The ratio of cpu time to simulation time,  given essentially by the 
number of floating point operations needed for one update of
the state,  depends linearly on the
number of layers $N_{c}$, but due to the quartic energy term 
for large systems quadratically on the 
number of vortices $N_{ab}$. Finite size effects become 
important and therefore large system sizes necessary when
correlation lengths grow large, which is unfortunately just 
in the regime of phase space where relaxation times also
grow large, namely at low $\alpha_{T}$.

\subsection{Effects of limited $N_{ab}$}
\label{sec:fsnab}
Most of our data have been taken using $N_{ab}=72$. For 
high temperatures finite size effects are negligible, 
because there is little in-plane order and
the range of crystalline correlations in any one layer
is much smaller than the spherical dimensions. The topological
disorder due to the 12 topological defects imposed by spherical
geometry \cite{dodgson} becomes at high temperatures negligible against the 
strong thermal disorder in the system. 
With decreasing $\alpha_{T}$ and 
growing $l_{ab}$, finite size effects due to finite $N_{ab}$
become important. For any one $\alpha_{T}$, the effects of
the limited number of vortices are the more severe
the  lower the coupling strength $|\alpha_{2T}\eta|$ and
the 2D effective temperature  $\alpha_{2T}$ are, because outside 
the continuum limit in-plane order increases with decreasing 
coupling strength, 
indicated by a smaller correlation length $l_{ab}$ (see Fig.\
\ref{fig:statlen}) and smaller $\beta_{A}$ (see Fig.\
\ref{fig:ba}). As a consequence of too small $N_{ab}$, we see a 
decrease in length scales of crystalline correlations as previously 
observed in 2D systems \cite{dodgson} as well as a decrease 
of correlations along the $c$-axis. 
We interpret agreement of measurements
taken with $N_{ab}=72$  and $N_{ab}=144$ as indication 
that with $N_{ab}=72$ we are already close to the thermodynamic 
limit. 

\subsection{Effects of limited $N_{c}$}
Finite size effects due to limited numbers of layers are 
for equilibrium measurements not a severe problem, because an 
increase of $N_{c}$ increases the
cpu time only linearly, and improves the rate of thermal
averaging by the same amount. Therefore the disadvantage  of 
increasing $N_c$ is only the increase in equilibration time as long 
as enough memory is available.
If $N_{c}/2$, the distance over which correlations 
decay, is not distinctly larger than the  correlation
length along the layers (see Sec.\ \ref{sec:opc}), we find that 
correlations both parallel and perpendicular to the $c$-axis are 
artificially enhanced and slower than exponential decay
behavior is observed, as visible in Fig.\ \ref{fig:cp_fsef}
for $N_c=30$. 
For equilibrium  measurements of length and time scales,
we always have  $N_{c}>10\times l_{c}/d$, which 
required on occasion using $N_{c}$ up
to 300 layers with 144 vortices per layer.

\subsection{Thermal averaging and initial relaxation}
\begin{figure}
\centerline{\epsfxsize=8cm\epsfbox{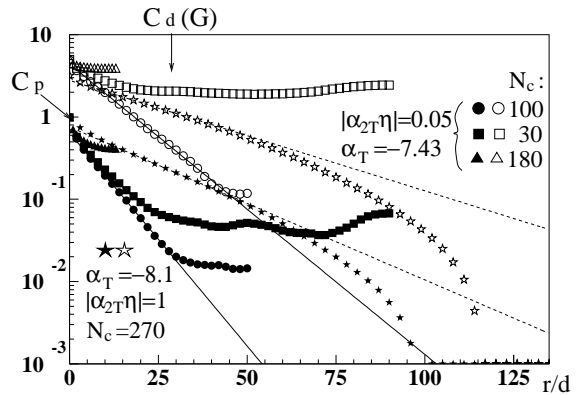}}
\caption{Typical measurements of static correlations along the $c$-axis.
Open symbols are density correlations,
filled symbols are phase correlations. Note for 
squares that correlations are enhanced by insufficient system size 
and for triangles the incomplete relaxation of the system.
Statistical noise can for large $r$ suggest slower than exponential 
decay (circles) as well as  faster than exponential decay (stars). 
For all systems $N_{ab}=72$.}
\label{fig:cp_fsef}
\end{figure}

Initial relaxation has proved a 
very difficult problem in our 3D simulations, where relaxation
times are so long that the entire simulation time is often
limited to only few times the longest relaxation time.
In principle, the relaxation can be started from an 
arbitrary state, where obvious choices are a
random state or a ground state. 
The relaxation from a ground state has the disadvantage
that it is very slow for low $\alpha_{T}$. On the other hand it is
fairly easy to judge how far the system has relaxed when 
started with the same state in every layer. Because our
system is always in a liquid state, a slower than exponential decay 
of correlations in the $c$-direction (which is in a sufficiently 
large system always removable by further equilibration)
is a reliable sign of insufficient equilibration.
A case of insufficient relaxation
can be seen in Fig.\ \ref{fig:cp_fsef} for $N_{c}=180$.
When using systems started from a ground state, they are 
always allowed to relax for longer than the  longest relaxation 
time $\tau$, in most cases longer than 5$\tau$. 
Relaxation from a random state has got the advantage of being 
faster than from a ground state. However, we found measurements 
from an insufficiently equilibrated random state often 
indistinguishable from equilibrium measurements at a higher
temperature, and therefore avoided starting 
equilibrium measurements from a random state.  

The most efficient way to obtain a well equilibrated 
system is to cool down or heat up a  
configuration obtained for a  similar temperature and coupling 
strength and identical system size. We used this method 
whenever such configurations were 
available. The cooling or heating rates in these cases are similar to
those used in our hysteresis measurements.
\vspace{-0.5cm}
\subsection{Loss of first order behavior}
\label{sec:fsef}
\begin{figure}
\centerline{\epsfxsize=9cm\epsfbox{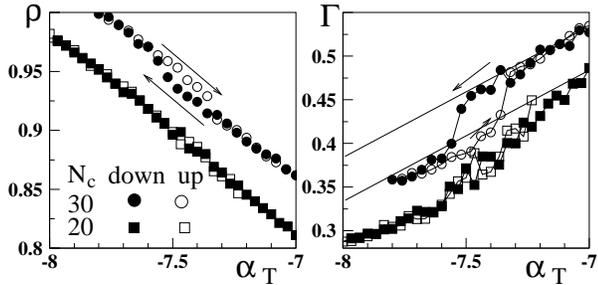}}
\caption[Loss of hysteresis with decreasing $N_{c}$]{Order parameter density 
$\rho$ and the measure of layer independence $\Gamma$
upon heating and cooling.
For insufficient number of layers, $\Gamma$ is reduced on the high
temperature
side of the transition and the first order behavior is lost.
For $N_{c}=20$ $\rho$ and  $\Gamma$ are offset by --0.05 
(equal levels marked by solid lines).
$N_{ab}=72$ and $|\alpha_{2T}\eta|=0.4$.
}
\label{fig:fsefc}
\end{figure}

The change of the hysteresis at the first order transition 
for decreasing numbers of layers can be seen in Fig.\
\ref{fig:fsefc}. The correlation length along the $c$-axis
(see Sec.\ \ref{sec:opc}) just above the transition is 
for $|\alpha_{2T}\eta|=0.4$ approximately
6 layers.  For $N_{c}=20$, correlations along
the $c$-axis are enhanced significantly just above the 
transition. The system does not remain in the high entropy, decoupled state,
and the hysteresis is lost. This effect is visible in Fig.\ 
\ref{fig:fsefc} in the reduction of the layer independence parameter 
$\Gamma$ (defined in Sec.\ \ref{sec:simfot}) just above the transition
in case of the smaller system
size. The loss of hysteresis with decreasing system size is rather
sudden. Further increase in system size beyond 5-6 times the
range of $c$-axis correlations affects the hysteresis measurements 
very little. 

In all cases the system sizes used for the hysteresis measurements
are more than five times the correlation length in $c$-direction
just above the transition.
The correlation lengths are for selected 
values of $|\alpha_{2T}\eta|$ known from equilibrium measurements in 
large systems and otherwise estimated by interpolation. 
We are aware that the system sizes used for the sweep
measurements are in most cases not distinctly larger than the 
range of correlations {\it below} the transition. However, the
observed size of the jumps does not seem much affected by this;
the discontinuities in the order parameter magnitude seen
in  equilibrium measurements in much larger systems for
selected $|\alpha_{2T}\eta|$ agree well with the results 
from sweep measurements. 
\vspace{-0.5cm}

A very important point to verify is that the loss of hysteresis at the
critical point is not an effect of insufficient numbers of layers.
We have made sure that the system sizes used near the critical point
are not only more than five times the correlation length,
but also that the loss of the transition at  $|\alpha_{2T}\eta|$=2.5 
occurs in a system that is
not only in terms of layers, but also in terms of the natural length
scale $\xi_{||}$, larger than the system for $|\alpha_{2T}\eta|$=2,
where a transition is still visible. 
For $|\alpha_{2T}\eta|$=2 and $N_{c}d/\xi_{||}=65$ we see a clear 
transition, while for $|\alpha_{2T}\eta|$=2.5 and  
$N_{c}d/\xi_{||}=80$ there is no sign of a transition in the 
range $-8.3<\alpha_{T}<-7.3$. 
\vspace{-0.5cm}
\begin{figure}
\centerline{\epsfxsize=9cm\epsfbox{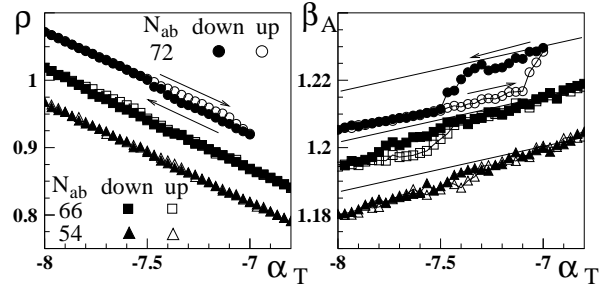}}
\caption[Loss of hysteresis with decreasing $N_{ab}$]{
Order parameter density 
$\rho$ and Abrikosov number $\beta_{A}$ upon heating and cooling.
For insufficient number
of vortices per layer, $\beta_{A}$ is increased on the low temperature
side of the transition and the first order behavior is lost.
For $N_{ab}=66$ and  $N_{ab}=54$, $\rho$ is offset by --0.05 and --0.1 
respectively and $\beta_{A}$  is offset by --0.015 and --0.03
respectively (equal levels marked by solid lines).
$N_{c}=12$ and $|\alpha_{2T}\eta|=0.05$.
}
\label{fig:fsefab}
\end{figure}

If the number of vortices per layer $N_{ab}$ is reduced,
the hysteresis at the transition first decreases and then
disappears. For the general reasons 
outlined in Sec.\ \ref{sec:fsnab} the effect  becomes 
stronger as $|\alpha_{2T} \eta|$ decreases.
Below some critical $N_{ab}$, which increases with decreasing 
$|\alpha_{2T}\eta|$, the in-plane order below the transition,
reflected in the Abrikosov Ratio $\beta_A$, is so much
affected by the spherical topology that the discontinuity
in $\beta_A$ vanishes and the transition disappears. 
This behavior is shown in Fig.\ \ref{fig:fsefab} for 
$|\alpha_{2T}\eta|=0.05$. The size of the discontinuities
decreases noticeably between $N_{ab}=72$ and  $N_{ab}=66$. 
For $N_{ab}=54$ the transition has disappeared. 
For $N_{ab}=72$, we still measure a transition  at 
$|\alpha_{2T}\eta|=0.02$, but not at  $|\alpha_{2T}\eta|=0.01$.
Thus the need of increasing $N_{ab}$ limits the exploration of the
phase diagram for very low $|\alpha_{2T}\eta|$.

We estimate that the  limitation of the number of vortices 
we study to $N_{ab}\!=\!72$
affects our hysteresis measurements for $|\alpha_{2T}\eta| \leq 0.05$.
We can detect size dependence in the location of 
the phase transition  for system sizes $N_{ab}=72$ and $N_{ab}=144$
only  for $|\alpha_{2T}\eta|<0.05$ (see the phase diagram in 
Fig.\ \ref{fig:alph}). The hysteresis measurements 
plotted in the same figure show that 
the size of the discontinuities at the phase transition does not
change noticeably between  $N_{ab}=72$ and $N_{ab}=144$
for $|\alpha_{2T}\eta|= 0.14$. 
For $|\alpha_{2T}\eta| \gg 0.14$, which applies to the region near 
the critical point, the system should at the phase transition 
be well simulated using $N_{ab}=72$. This is confirmed by 
the agreement in the 
magnetization discontinuity between $N_{ab}=72$ and $N_{ab}=144$
in Fig.\ \ref{fig:btdiamag}(b) for $T$=89.3.

\end{multicols}

\begin{references}
\bibitem{footnote}
The authors of Ref. \protect \cite{nguyen_new}
do not report clear evidence for a first order transition. However,
we believe that a peak in the numerical specific heat data 
of Ref. \protect \cite{nguyen_new} at the transition 
may indicate a first order transition.
\end{references}
\end{document}